\documentclass[aps,showpacs,prx,superscriptaddress,preprintnumbers,amsmath,amssymb,twocolumn]{revtex4-2}
\usepackage[utf8]{inputenc}
\usepackage[T1]{fontenc}
\usepackage{graphicx}
\usepackage{dcolumn}
\usepackage{appendix}
\usepackage{amsmath}
\usepackage{bm}
\usepackage[dvipsnames]{xcolor}
\usepackage{braket}
\usepackage{bbold}
\usepackage{mathtools}
\usepackage{float}

\usepackage{subfigure}
\usepackage{bbm}
\usepackage{enumerate}

\usepackage[
  bookmarks=true,
  colorlinks,
  linkcolor=black,
  urlcolor=black,
  citecolor=black,
  plainpages=false,
  pdfpagelabels,
  final,
  breaklinks=true
]{hyperref}

\usepackage{titlesec}

\usepackage{array}

\usepackage{physics}

\definecolor{blue-violet}{rgb}{0.54, 0.17, 0.89}

\begin{document}
\allowdisplaybreaks
\title{Device-Independent Quantum Key Distribution beyond qubits}

\author{Javier Rivera-Dean}
\email{javier.rivera@icfo.eu}
\thanks{These authors contributed equally to this work}
\affiliation{ICFO -- Institut de Ciencies Fotoniques, The Barcelona Institute of Science and Technology, 08860 Castelldefels (Barcelona)}

\author{Anna Steffinlongo}
\thanks{These authors contributed equally to this work}
\affiliation{ICFO -- Institut de Ciencies Fotoniques, The Barcelona Institute of Science and Technology, 08860 Castelldefels (Barcelona)}

\author{Neil Parker-Sánchez}
\thanks{These authors contributed equally to this work}
\affiliation{ICFO -- Institut de Ciencies Fotoniques, The Barcelona Institute of Science and Technology, 08860 Castelldefels (Barcelona)}

\author{Antonio Acín}
\affiliation{ICFO -- Institut de Ciencies Fotoniques, The Barcelona Institute of Science and Technology, 08860 Castelldefels (Barcelona)}
\affiliation{ICREA -- Instituci\'o Catalana de Recerca i Estudis Avan\c cats, Llu\'{\i}s Companys 23, 08010 Barcelona, Spain}

\author{Enky Oudot}
\email{enky.oudot@icfo.eu}
\affiliation{ICFO -- Institut de Ciencies Fotoniques, The Barcelona Institute of Science and Technology, 08860 Castelldefels (Barcelona)}

\date{\today}
\begin{abstract}
Device-Independent Quantum Key Distribution (DIQKD) aims to generate secret keys between two parties without relying on trust in their employed devices, imposing strict noise constraints for key generation. This study explores the resilience of high-dimensional quantum systems in DIQKD, focusing on a comparison between qubits and qutrits. Lower bounds on achievable key rates are investigated through numerical optimization, while upper bounds are evaluated using the Convex-Combination attack, which has been further extended to account for arbitrary dimensions. The observed difference between these bounds provides insights into noise thresholds and potential enhancements in DIQKD scenarios, prompting debate on the merit of increased dimensions given the associated experimental efforts required.

\end{abstract}
\maketitle

\section{INTRODUCTION}
Quantum Key Distribution (QKD) stands as one of the most promising and successful applications stemming from the \emph{second quantum revolution}~\cite{Benyoucef_book_ch2}, aimed at leveraging and harnessing the properties of quantum mechanics toward novel technological advancements. In QKD protocols, the security of the established key among two or more parties relies on both the principles of quantum physics and the precise description of the experimental apparatus. However, minor deviations from these exact specifications of the used protocol can enable eavesdroppers to compromise its security~\cite{makarov_effects_2006,qi_time-shift_2007,zhao_quantum_2008,makarov_controlling_2009,lydersen_hacking_2010,gerhardt_full-field_2011}. 

In this context, Device-Independent Quantum Key Distribution (DIQKD) seeks to overcome potential vulnerabilities associated with the trustworthiness of the quantum devices employed for communication~\cite{Acin07}. DIQKD specifically shifts its emphasis from trusting the internal functionalities of quantum devices to relying solely on observed correlations between measurements conducted by distant parties. In this regard, the price to pay for removing the requirements for a physical description of the measurement apparatus is the observation of a substantial violation in a Bell test~\cite{Acin07}, thus tolerating low levels of noise~\cite{nadlinger_experimental_2022,zhangExperimentalDeviceindependentQuantum2022,China_exp_2022,gonzalez-ruiz_device_2022}. Notably, the security of such protocols has been
successfully demonstrated, even in scenarios allowing an eavesdropper to execute general attacks, see e.g. Ref.~\cite{Friedman19}.

The security of DIQKD protocols relies on the violation of a Bell inequality~\cite{Acin07}. Intuitively, for certain Bell inequalities such as the Clauser-Horne-Shimony-Holt (CHSH) inequality~\cite{Clauser69}, maximal violation observed by two parties indicates their sharing of a maximally entangled two-qubit state~\cite{mayers_self_2004, Kaniewski16}, rendering them uncorrelated with any third party. However, achieving the maximum violation of such an inequality is hindered by inevitable noise, prompting extensive research efforts to enhance the noise robustness of security proofs~\cite{ho_noisy_2020,brown2023deviceindependent}. Simultaneously, minimum noise requirements have been established through the deliberate design of potential eavesdropping attacks on a given protocol~\cite{farkas_bell_2021}. The discrepancy between these requirements has significantly narrowed, allowing for minimal potential improvement~\cite{gonzalez-ruiz_device_2022,Lukanowski:23}, especially in scenarios where the shared state is encoded using qubits. Notably, recent successful DIQKD experiments employing qubit-encoded shared states have been recently conducted~\cite{nadlinger_experimental_2022,zhangExperimentalDeviceindependentQuantum2022}.

To scale up DIQKD to medium or long distances, either experiments need to meet the requirements derived from security proofs or the scenario itself has to change. One potential avenue involves augmenting the number of inputs and outputs in the protocol, or increasing the shared quantum system's dimensionality between Alice and Bob. This could present a viable approach for DIQKD protocols, particularly considering that maximally entangled states can self-test across arbitrary local dimensions~\cite{Sarkar21}. Furthermore, compared to qubits, utilizing higher-dimensional systems has been shown to increase the noise robustness of violations of Bell inequalities~\cite{collinsBellInequalitiesArbitrarily2002,salavrakosBellInequalitiesTailored2017}, of the security of device-dependent QKD protocols~\cite{Cerf_qudit_QKD_2002}, and also allows for device-independent extraction of a greater number of random bits~\cite{randomness_qutrit}.

This study investigates the impact of increased inputs, outputs, and shared system dimensions between Alice and Bob on the noise requirements in DIQKD protocols. We analyze the security of these protocols, aiming to establish both upper~\cite{brown2023deviceindependent} and lower bounds~\cite{farkas_bell_2021} for noise requirements, ensuring the generation of secure keys between two parties. Specifically, our focus lies in comparing qubits and qutrits, although an extension of the lower bounds for noise requirements to encompass arbitrary dimensions is also presented.

 Before proceeding, it is worth clarifying that in DIQKD security proofs, Hilbert space dimension does not play any role, but the cardinality of the measurement outputs $s_A$ and $s_B$, here taken equal to $d$, is the relevant parameter. However, when implementing a DIQKD protocol, this parameter can be associated to the Hilbert space dimension of the measured quantum systems when local projective measurements are applied to them, or to the extended Hilbert spaces defined by local non-projective POVMs, consisting of projective measurements on the systems and ancillas.

\section{SCENARIO}
In QKD scenarios, two spatially separated and trustworthy parties, hereupon identified as Alice and Bob, aim to establish a secure secret key for communication between them. Meanwhile, a potential adversary, referred to as Eve in the following, attempts to eavesdrop on their communication in pursuit of accessing information regarding the shared key. 

\begin{figure*}
    \centering
    \includegraphics[width=1\textwidth,trim={0 17cm 0 0}]{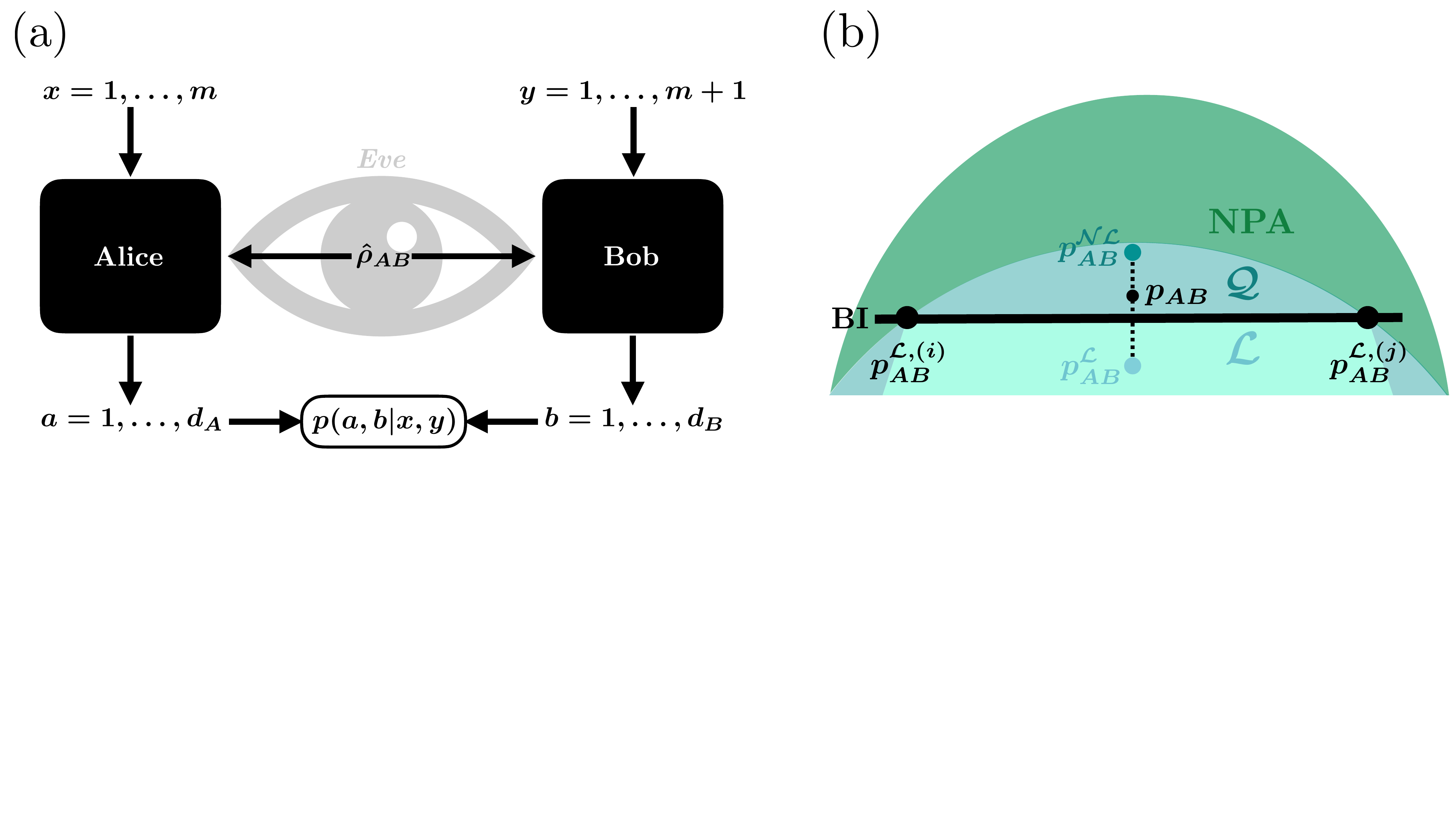}
    \caption{In (a), a graphical representation of a typical DIQKD protocol is depicted. Alice and Bob input measurement parameters $x$ and $y$ into their respective black boxes, resulting in outcomes $a$ and $b$ in return. This entire process establishes a set of observed correlations $\{p(a,b\vert x,y)\}$, defining the likelihood of obtaining outcomes $a$ and $b$ given the introduction of measurements $x$ and $y$. In DIQKD, the protocol's security against an eavesdropper, Eve, is determined based on these correlations established between Alice and Bob. In panel (b), an illustrative scheme depicts the methods employed to compute upper and lower bounds on the key rate. For the analysis of lower bounds, Eve is permitted to execute attacks utilizing correlations beyond the quantum set $\mathcal{Q}$, illustrated by the ${\text{NPA}}$ set. For the upper bounds, the CC attack is employed. In this scenario, Eve selectively sends $p^{\mathcal{L}}_{AB}$ or $p^{\mathcal{NL}}_{AB}$, with the constraint that their linear combination reproduces the observed correlations $p_{AB}$.}
    \label{Fig:scheme:setup}
\end{figure*}

In this scenario, the joint state describing this system is generally represented as a $(d_A \times d_B \times d_E)$-dimensional tripartite quantum state $\hat{\rho}_{ABE}$ acting on $\mathcal{H}_A\otimes\mathcal{H}_B\otimes\mathcal{H}_E$, where $\mathcal{H}_i$ denotes the Hilbert space of party $i$ ($i \in \{A,B,E\}$), satisfying $\dim(\mathcal{H}_i) = d_i$. Thus, in a first step towards generating a secret key, Alice and Bob manipulate physical systems that perform local operations on their respective share of $\hat{\rho}_{ABE}$, producing outputs utilized later in key generation. Specifically, by randomly selecting classical inputs, labeled here as $x \in \{1, \dots, m\}$ for Alice and $y\in \{1, \dots m+1\}$ for Bob, the systems yield outputs $a\in\{1,\dots,s_A\}$ and $b\in\{1,\dots,s_B\}$, respectively. In the following, we consider $s_A=s_B=d_A = d_B = d$. The measurements performed on these systems can be defined by sets of Positive Operator-Valued Measures (POVMs), denoted as $\{\{\hat{\Pi}_{a|x}\}_a\}_x$ for Alice and $\{\{\hat{\Pi}_{b|y}\}_b\}_y$ for Bob. These measurements, along with the quantum state $\hat{\rho}_{ABE}$, establish joint conditional probability distributions ${p(a,b\vert x,y)}$, signifying the probability of obtaining outputs $a$ and $b$ given the implementation of measurement inputs $x$ and $y$. Employing the Born rule, $p(a,b\vert x,y)$ can be expressed as
\begin{equation}\label{Eq:joint:prob}
    p(a,b\vert x,y)=\text{Tr} \big[ \hat{\rho}_{ABE}(\hat{\Pi}_{a|x}\otimes\hat{\Pi}_{b|y}\otimes\mathbb{1} )\big].
\end{equation}
The set of quantum correlations or quantum set $\mathcal Q$ is defined by those joint conditional probability distributions ${p(a,b\vert x,y)}$ that can be written in the form of Eq.~\eqref{Eq:joint:prob}.

In the realm of DIQKD, the primary goal is to study the security of the protocol based on the set of correlations $p_{AB} \coloneqq\{p(a,b\vert x,y)\}$ describing Alice's and Bob's outputs statistics after performing the aforementioned experiment a given number of rounds $n$, without relying on trust in the utilized measurements nor the shared state. Essentially, they treat their measurement devices as black boxes, as pictorially presented in Fig.~\ref{Fig:scheme:setup}~(a). Specifically, the initial $m$ measurements performed by both parties aim to validate a Bell inequality violation, ensuring the existence of nonlocal correlations shared between Alice and Bob. Conversely, the outputs acquired from inputs $x^*=m$ and $y^*=m+1$ are utilized for the raw key generation.

The security of the protocol is quantified by the key rate $r$, indicating the number of secure bits generated by Alice and Bob per protocol round. 
Imperfections in Alice's and Bob's measurement devices, which could potentially stem from the presence of an eavesdropper seeking information about the key, lower the value of $r$. A protocol is deemed secure whenever $r > 0$. The threshold case $r=0$ provides the conditions that must be satisfied to ensure the generation of a secure key.

The key rate $r$ for one-way protocol communication is expressed as the difference between an error-correction (EC) term, which determines the fraction of bits Alice has to publicly communicate to Bob in order to correct any potential mismatch between their raw keys; and the privacy amplification (PA) term, representing the fraction of bits Alice has to compress in order to ensure that Eve has zero knowledge of the resulting key~\cite{renner_information-theoretic_2005}. In this work, we study upper bounds $r_{\text{ub}}$ and lower bounds $r_{\text{lb}}$ on the key rate, that is
\begin{equation}
    r_{\text{ub}} \geq r \geq r_{\text{lb}},
\end{equation}
obtained by bounding the PA term in two different ways, pictorially represented in Fig.~\ref{Fig:scheme:setup}~(b).

One approach to establishing a lower bound on the key rate involves overestimating Eve's knowledge regarding Alice's outcomes, that is, lower bounding the PA-term. This is achieved by allowing Eve to execute attacks that surpass correlations within the quantum set $\mathcal{Q}$. The Navascués-Pironio-Acín (NPA) hierarchy facilitates this overestimation~\cite{navascues_bounding_2007,navascues_convergent_2008}, leading to the NPA set presented in Fig.~\ref{Fig:scheme:setup}~(b). Conversely, an upper bound can be derived by introducing specific quantum strategies followed by Eve to predict Alice's outcome. We examine the Convex-Combination (CC) attack for this purpose~\cite{farkas_bell_2021} which, in certain scenarios, has demonstrated close alignment with state-of-the-art techniques used to compute lower bounds~\cite{Lukanowski:23}. In the CC attack, Eve mimics the correlations observed by Alice and Bob, i.e., $p_{AB}$, by randomly alternating between correlations compatible with the local-hidden-variable-model~\cite{brunnerBellNonlocality2014} or non-local correlations, represented in Fig.~\ref{Fig:scheme:setup}~(b) as $p^{\mathcal{L}}_{AB}$ and $p^{\mathcal{NL}}_{AB}$ respectively.

The difference between these two bounds offers insights into noise requirements and potential enhancements achievable in a given DIQKD scenario. This study delves into these perspectives, exploring the impact of increasing the dimension $d$ and varying number of measurement choices $m$ on the key rate. Specifically, our focus for the lower bounds resides within the parameter space where $d,m \in \{2,3\}$, whereas analytical upper bounds enable us to increase $d$ further.

\section{BOUNDING THE KEY RATE}\label{Sec:Bounds}

In this section, we provide an overview of the methods used to compute both the lower and upper bounds of the key rate, offering a conceptual understanding without delving into specific details. More comprehensive information on calculations and methodology is available in the supplementary material (SM).

\subsection{Lower bounds}\label{sec:LB}
In one-way scenarios, where the parties publicly communicate in one direction, say, from Alice to Bob, a lower bound to the key rate in the asymptotic regime $n\to\infty$ is provided by the commonly known Devetak-Winter (DW) bound~\cite{Devetak_2005}
\begin{equation}\label{Eq:DW_bound}
    r_{\text{DW}}=H(A\vert x=x^*, E)-H(A\vert B, x=x^*, y=y^*),
\end{equation}
where $H(A\vert x=x^*,E)$ represents the conditional entropy between Alice's outcome when measuring $x^*$ and Eve, and $H(A\vert B,x^*,y^*)$ denotes the conditional entropy between Alice's and Bob's outcomes when performing measurements $x^*$ and $y^*$, respectively. Notably, for the latter term, the additional measurement $y^*$ performed by Bob can be optimized with the objective of minimizing its value.

Given that the conditional probabilities in Eq.~\eqref{Eq:joint:prob} only depend on a particular implementation of the protocol, the computation of $H(A\vert B, x=x^*, y=y^*)$ becomes straightforward. However, the same does not hold for $H(A\vert x=x^*, E)$, which we bound numerically instead. In this context, we adopt two distinct approaches to lower bound the entropy: the min-entropy and a convergent hierarchy to the von Neumann entropy. Regarding the former, it is given by~\cite{konig_operational_2009,acin_randomness_2012}
\begin{equation}\label{Eq:min_entrpy}
    H_{\text{min}}(A|x=x^*, E)=-\log_d G(A|x=x^*, E),
\end{equation}
where $G(A|x=x^*, E)$ is the guessing probability~\cite{acin_randomness_2012} (see SM~\ref{App:Optimization}), that is, Eve's probability of guessing Alice's outcome when performing measurement $x^*$. Alternatively, in Ref.~\cite{brown2023deviceindependent}, the authors derived a convergent series of lower bounds on $H(A\vert x=x^*, E)$ given by
\begin{equation}\label{Eq:BFF:bound}
    \begin{aligned}
    \Tilde{H}^{(M)}(A\vert x=x^*,E) =
        c_M 
            + \sum_{i=1}^{M-1} \dfrac{w_i}{t_i \ln{d}} 
                \sum^d_{a=1} f\big(t_i, \hat{\Pi}_{a\vert x^*}\big),
    \end{aligned}
\end{equation}
where $f(t_i, \hat{\Pi}_{a\vert x^*})$ is a function to be optimized over Eve's measurements subjected to a set of linear constraints. In this expression, $w_i$ and $t_i$ are the $i$th Gauss-Radau quadratures, with $M$ being the total number of nodes, such that increasing values of $M$ provide tighter bounds on $H(A\vert x=x^*,E)$ in Eq.~\eqref{Eq:DW_bound}  (for details we refer the reader to SM~\ref{App:BFF}). 

While to our knowledge there is no direct link between Eqs.~\eqref{Eq:min_entrpy} and \eqref{Eq:BFF:bound}, in practice it is obtained that for sufficiently large enough values of $M$ ($M \geq 8$), Eq.~\eqref{Eq:BFF:bound} yields superior bounds on the key rate compared to Eq.~\eqref{Eq:min_entrpy}. However, optimizing Eq.~\eqref{Eq:BFF:bound} given certain parametrized POVM sets and a shared quantum state between Alice and Bob is a significantly intricate task~\cite{brown2023deviceindependent,gonzalez-ruiz_device_2022}. Particularly in the scenarios examined here, coupled with the available computational resources, this optimization becomes nearly impractical. In such circumstances, the utilization of the min-entropy, which is computationally more manageable compared to Eq.~\eqref{Eq:BFF:bound}, becomes notably advantageous.

\begin{figure}
    \centering
    \includegraphics[width=1\columnwidth]{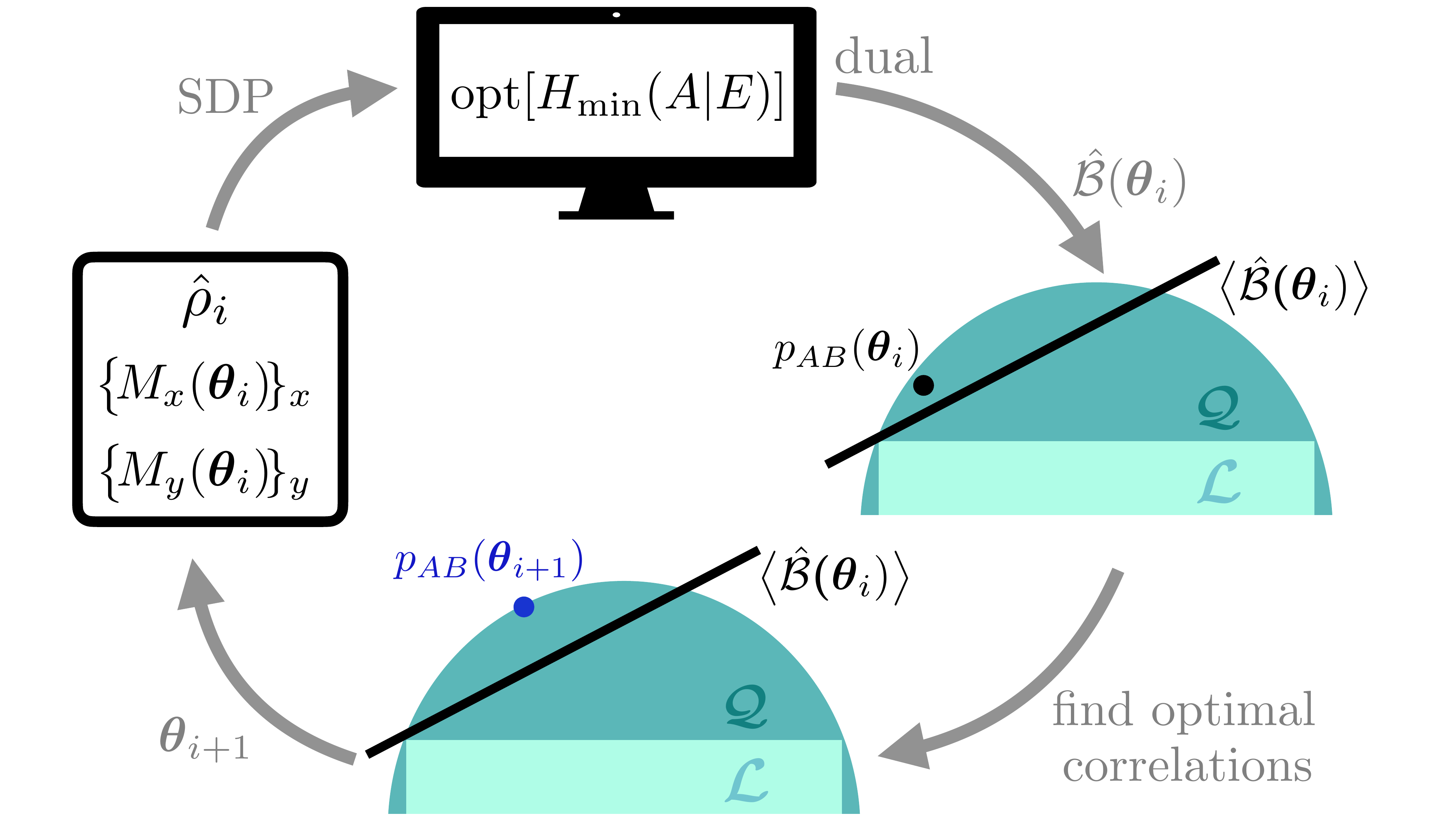}
    \caption{A visual representation of the methodology employed to optimize Eq.~\eqref{Eq:min_entrpy} with respect to the parameters $\boldsymbol{\theta}$. Given a state and a series of measurements conducted by Alice and Bob, these establish a set of constraints enabling the optimization of $H_{\text{min}}(A|E)$ via Semi-Definite Programming (SDP). The dual of this SDP provides us with a Bell inequality, which is optimized by means of local optimization methods with respect to $\boldsymbol{\theta}$ to derive a new state and a set of measurements applicable in subsequent steps.}
    \label{fig:scheme:optimization}
\end{figure}

Consequently, by considering a parameterization of the measurement operators $\{M_x(\boldsymbol{\theta})\}_x$ and $\{M_y(\boldsymbol{\theta})\}_y$ (further described in SM~\ref{App:measurements}), the method employed here to derive the lower bound on the key rate comprises two primary steps: (1) a semi-definite optimization, and (2) a local optimization involving the parameters $\boldsymbol{\theta}$. A pictorial representation of these steps is illustrated in Fig.~\ref{fig:scheme:optimization}, with a detail step by step explanation available in SM~\ref{App:Optimization}.

At the $i$th step, the optimization commences with a quantum state $\hat{\rho}_i$, expressed in the form
\begin{equation}
\label{eqstateV}
    \hat{\rho_i}
        = V \dyad{\psi_i}
            + \dfrac{1-V}{d} \mathbbm{1},
\end{equation}
with $V$ referred to as the visibility, and the parameters $\boldsymbol{\theta}_i$ yielding the measurement sets $\{M_x(\boldsymbol{\theta}_i)\}_x$ and $\{M_y(\boldsymbol{\theta}_i)\}_y$ for Alice and Bob, respectively. These quantities provide a series of linear constraints, enabling the computation of the optimal value for $H_\text{min}(A\vert x=x^*,E)$ through Semi-definite Programming (SDP) methods~\cite{BoydCH4}. Here, we take into advantage the fact that SDPs can be formulated in two equivalent ways: as a minimization involving a certain objective function, in our case Eq.~\eqref{Eq:min_entrpy}, known as the primal problem; or as a maximization over the set of constraints, termed the dual problem. Consequently, the dual of the optimized SDP yields a Bell inequality from which we construct a Bell operator $\hat{\mathcal{B}}(\boldsymbol{\theta}_i)$ that, by definition, depends on $\hat{\rho}_i$ and the measurement settings $\{M_x(\boldsymbol{\theta}_i)\}_x$ and $\{M_y(\boldsymbol{\theta}_i)\}_y$. This Bell operator defines a Bell inequality $\langle\hat{\mathcal{B}}(\boldsymbol{\theta}_i)\rangle$, which is violated by the correlations obtained from $\hat{\rho}_i$, $\{M_x(\boldsymbol{\theta}_i)\}_x$ and $\{M_y(\boldsymbol{\theta}_i)\}_y$, represented as $p_{AB}(\boldsymbol{\theta}_i) \equiv \{p(a,b|x(\boldsymbol{\theta}_i),y(\boldsymbol{\theta}_i))\}$ in Fig.~\ref{fig:scheme:optimization}. Consequently, optimizing the maximal violation of $\langle\hat{\mathcal{B}}(\boldsymbol{\theta}_i)\rangle$ via local optimization methods leads to a new state $\hat{\rho}_{i+1}$  satisfying \eqref{eqstateV}, and an optimal set of parameters $\boldsymbol{\theta}_{i+1}$ that can be utilized in the SDP optimization for $H_{\text{min}}(A|x=x^*,E)$. These outlined steps are iteratively optimized until convergence of $H_\text{min}(A\vert x=x^*,E)$ is attained.

Upon reaching convergence, the additional measurement settings employed by Bob to compute the key, denoted as $M_{y^*}(\boldsymbol{\theta}_i)$, underwent optimization aimed at minimizing the error correction term $H(A\vert B, x=x^*, y=y^*)$. Subsequently, the optimal parameters were utilized to calculate $\tilde{H}^{(M)}(A\vert x=x^*,E)$ in Eq.~\eqref{Eq:BFF:bound}, with $M=16$, and determine the key rate.

\subsection{Upper bounds}\label{sec:UB}

In order to construct an upper bound on the one-way key rate, we generalize to dimension $d$ the approach followed by \Citet{Lukanowski:23}, based on the CC attacks originally proposed by \Citet{farkas_bell_2021}. These are individual attacks in which Eve's strategy is to distribute, in each round, either local bipartite correlations $p_{AB}^\mathcal{L}(a,b|x,y)$ with probability $q^\mathcal{L}$, or a non-local one $p_{AB}^\mathcal{NL}(a,b|x,y)$ with probability $q^\mathcal{NL}=1-q^\mathcal{L}$. To reproduce the observed correlations, these must satisfy
\begin{multline} \label{eq:cc_attack}
    q^\mathcal{L} p_{AB}^\mathcal{L} (a,b|x,y) + q^\mathcal{NL} p_{AB}^\mathcal{NL} (a,b|x,y) = p_{AB} (a,b|x,y) \\
    \forall a,b,x,y.
\end{multline}
Since local correlations can be decomposed as a convex combination of deterministic strategies, that is $p_{AB}^\mathcal{L} = \sum_i \gamma_i p_{AB}^{\mathcal{L},(i)}$ with $\gamma_i\in[0,1]$ $\forall i$, Eve can distribute the deterministic strategy $p_{AB}^{\mathcal{L},(i)}$ in each round with probability $q_i^\mathcal{L} = \gamma_i q^\mathcal{L}$. By keeping track of the distributed deterministic strategy, Eve has perfect knowledge of Alice and Bob's outcomes for the key settings $x^*$ and $y^*$ in each local round. On the contrary, we make the overpessimistic assumption that Eve has no knowledge of their outcomes in the non-local rounds. 

For a particular individual attack, if Alice does not perform any preprocessing, the following expression provides an upper bound on the asymptotic key rate with one-way error correction
\begin{multline} \label{eq:upper_bound}
 r_{1-\textup{way}} (A \to B) \le H(A|x=x^*, E)- \\ H(A|B, x=x^*, y=y^*) =: r_\textup{ub}.
\end{multline}
Here, $H(A|x=x^*, E)$ is the PA-term and $H(A|B, x=x^*, y=y^*)$ is the EC-term. Henceforth, we omit any reference to the measurement settings, which we take to be the key settings.

In order to find the tightest possible upper bound, we must optimize the CC attack. In other words, we must maximize the knowledge gained by Eve. Once the observed correlations $p_{AB}$ and the non-local correlations used by Eve $p_{AB}^\mathcal{NL}$ are fixed, this corresponds to finding the local correlations that satisfies Eq.~\eqref{eq:cc_attack} and maximizes $q^\mathcal{L}$. This can be expressed in terms of the following linear optimization problem
\begin{equation} \label{eq:linear_program}
    \begin{aligned}
        & \textup{Find a vector} & &\mathbf{q} \coloneqq (\mathbf{q^\mathcal{L}}, q^\mathcal{NL}) \\
        & \textup{that maximizes} & &(1,1,\dots,1,0) \cdot \mathbf{q} \\
        & \textup{subject to} & &(1,1,\dots,1) \cdot \mathbf{q} = 1 \\
        &  & &\mathbf{0} \le \mathbf{q} \le \mathbf{1} \\
        &  & & \mathbf{q} \cdot (\mathbf{p_{AB}^\mathcal{L}}, p_{AB}^\mathcal{NL}) = p_{AB} 
    \end{aligned}
\end{equation}
where $\mathbf{p_{AB}^\mathcal{L}} = \{ p_{AB}^{\mathcal{L},(i)} \}_i$ is the set of all local deterministic strategies, $p_{AB}^\mathcal{NL}$ is the chosen non-local correlation, and $p_{AB}$ is the observed correlation.

In this work we assume that the non-local correlations used by Eve is the same as the ideal, noise-free, correlations Alice and Bob intend to share. We do this in order to find an upper bound on the key rate in the full range of visibilities $V \in[0,1]$. If Eve were to use non-local correlations different from Alice and Bob's, then in the limit of $V\to 1$ the CC attack would not be possible, since Eve would always have to distribute the same non-local correlations in order to match the observed correlations. Therefore, in the finite visibility scenario, the probabilities observed by Alice and Bob are
\begin{equation} \label{eq:finite_visibility}
    p_{AB} (a,b|x,y) = V p_{AB}^{\mathcal{NL}}(a,b|x,y) + \frac{1-V}{d^2}.
\end{equation}

In particular, we consider two non-local correlations, which are the ones that allow for maximal quantum violation of the inequalities introduced by \Citet{salavrakosBellInequalitiesTailored2017} and \citet{collinsBellInequalitiesArbitrarily2002}, respectively, the latter referred to as the CGLMP-inequality. We observe that these inequalities provide key rates that are in very good agreement, up to some mild differences, to those obtained with the optimization method depicted in Fig.~\ref{fig:scheme:optimization} (we refer the reader to SM~\ref{App:Optimization}). Furthermore, Salavrakos' inequality is chosen because its maximal violation provides a perfect secret $d$-value key. This is because it self-tests the maximally entangled state $\ket{\psi_0} = (1/\sqrt{d})\sum_{q=1}^{d}\ket{qq}$, and the associated optimal measurements, which we refer to as the CGLMP-optimal measurements \citep{salavrakosBellInequalitiesTailored2017}. These measurements also lead to the maximal violation of the CGLMP-inequality by this state \citep{collinsBellInequalitiesArbitrarily2002}. However, a larger CGLMP violation can be attained by another, non-maximally entangled, state, which we refer to as the CGLMP state. In fact, the maximal quantum violation of the CGLMP inequality obtained by the CGLMP state defines optimal correlations in terms of noise robustness~\cite{acinQuantumNonlocalityTwo2002}. 


\subsubsection{Maximally entangled state}

The CC attack can be optimized analytically if we choose as non-local term the correlations maximally violating Salavrakos' inequality, obtained by measuring a maximally entangled state. By substituting Eq.~\eqref{eq:finite_visibility} into Eq.~\eqref{eq:cc_attack} we can write
\begin{equation} \label{eq:prob_loc}
        p_{AB}^\mathcal{L} (a,b|x,y) = \tilde{V} p_{AB}^{\mathcal{NL}}(a,b|x,y) + \frac{1-\tilde{V}}{d^2}
\end{equation}
where $\tilde{V} \coloneqq \left(V-(1-q^\mathcal{L})\right)/q^\mathcal{L}$. Therefore, maximising $q^\mathcal{L}$  corresponds to maximising $\tilde{V}$ such that $p_{AB}^\mathcal{L} (a,b|x,y)$ is local. The result of this maximization is the \emph{local visibility} $V^\mathcal{L}$. Hence, the maximal local weight is
\begin{equation} \label{eq:qL}
    q^\mathcal{L} = \begin{cases}
    \dfrac{1-V}{1-V^\mathcal{L}} \quad &\textup{if }V\ge V^\mathcal{L} \\
    1 &\textup{otherwise}
    \end{cases}.
\end{equation}
To detect the nonlocality of the resulting correlations, we can use the CGLMP-inequality, since this inequality is tight, which means it coincides with a facet of the local polytope \citep{DBLP:journals/qic/Masanes03}. This inequality is expressed as \citep{collinsBellInequalitiesArbitrarily2002}
\begin{equation} \label{eq:CGLMP}
    \begin{aligned}
    I_d &= \sum_{k=0}^{[d / 2]-1}\left(1-\frac{2 k}{d-1}\right) \Big\{ p(A_{1}=B_{1}+ k) \\
    &\quad +p(B_{1}=A_{2}+k+1 )+p(A_{2}=B_{2}+k) \\
    &\quad+p(B_{2}=A_{1}+k) -p(A_{1}=B_{1}-k-1) \\
    &\quad-p(B_{1}=A_{2}-k) -p(A_{2}=B_{2}-k-1) \\
    &\quad -p(B_{2}=A_{1}-k-1) \Big\} \le 2 \eqqcolon C_b .
    \end{aligned}
\end{equation}
The maximum value of $\tilde{V}$ is the ratio between the local bound $C_b$ and the maximum violation of the CGLMP-inequality by the maximally entangled state $V^\mathcal{L} = C_b/I_d^\textup{max}$  (see SM~\ref{appendix:analytical_derivation} for details).
This allows us to determine the maximum local weight using Eq.~\eqref{eq:qL}. The conditional entropy $H(A|E)$ is $1$ for the non-local rounds and $0$ for the local rounds. Therefore $H(A|E) = 1- q^\mathcal{L}$.

Computing the conditional Shannon entropy for \eqref{eq:finite_visibility} yields the EC-term
\begin{multline}
        H(A|B) = - \frac{1+(d-1)V}{d}\log_d \left(1+(d-1)V \right) \\ \qquad- \frac{(d-1)(1-V)}{d} \log_d\left( 1-V \right) + 1.
\end{multline}
By subtracting these two terms, we get the following upper bound on the key rate:
\begin{equation}\label{Eq:r:up}
    \begin{aligned}
    r_{\text{ub}}
        &= \frac{1+(d-1)V}{d}\log_d \left(1+(d-1)V \right)
        \\&\quad+ \frac{(d-1)(1-V)}{d} \log_d\left( 1-V \right)
        -  \frac{1-V}{1-2/I_d^\textup{max}}.
    \end{aligned}
\end{equation}



\subsubsection{CGLMP state}

In the case of the CGLMP state, we use linear programming to solve the problem defined in Eq.~\eqref{eq:linear_program} for a given visibility in order to find the local weight and the corresponding upper bound on the key rate. To do this, we first need to find $p_{AB}^\mathcal{NL}$. For this, we first define the Bell operator $\hat{\mathcal{B}}_d$ corresponding to the CGLMP-inequality defined in Eq.~\eqref{eq:CGLMP}, where we use a measurement parameterization of the measurement operators which achieve the maximum quantum violation following \cite{salavrakosBellInequalitiesTailored2017, barrett_maximally_2006}. We then optimize this Bell operator to find the optimal state and measurements. We use this state as the non-local state and find $p_{AB}^\mathcal{NL}$.

\begin{figure}
    \centering
    \includegraphics[width=\columnwidth]{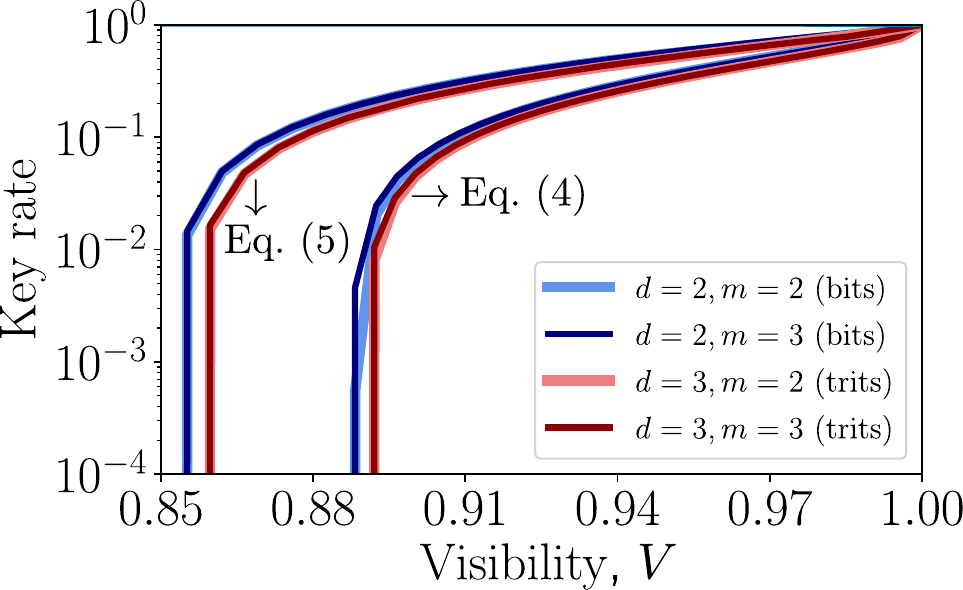}
    \caption{Lower bounds on the key rate attained when using the min-entropy~\eqref{Eq:min_entrpy} and the lower bound~\eqref{Eq:BFF:bound}, the latter when setting $M=16$. Four different scenarios are considered depending on the values of $d$ ($d=2$ in blue and $d=3$ in red) and $m$ ($m=2$ with light colors and $m=3$ with dark colors). Notably, the case of $d=2$, $m=2$ retrieves the critical visibilities found within the literature (e.g.,~Ref.~\cite{gonzalez-ruiz_device_2022}).}
    \label{Fig:lower:bounds}
\end{figure}

\section{RESULTS}
In this section, we present the results derived through the methodology outlined in Section~\ref{Sec:Bounds}. The section is structured into two subsections. The first subsection examines both lower and upper bounds on the key rate in relation to the visibility parameter $V$, focusing on cases where $d \in \{2,3\}$. This limitation primarily stems from computational constraints: optimization, as depicted in Fig.~\ref{fig:scheme:optimization}, becomes unfeasible due to memory limitations beyond these values. Consequently, obtaining analytical expressions for the upper bounds allows us to discern the impact of these visibility requirements for $d>3$, which is studied in the second subsection. Hereupon, the results for $d=3$ are shown in units of trits, while those for $d=2$ in units of bits.

\subsection{\boldmath Analysis of key rate bounds with respect to visibility for $d\in \{2,3\}$}

In both the analyses of lower and upper bounds presented in Sec.~\ref{Sec:Bounds}, a dichotomy emerged concerning the bounding of the conditional entropy $H(A\vert x=x^*,E)$. The former analysis raised the question of utilizing either Eq.~\eqref{Eq:min_entrpy} or Eq.~\eqref{Eq:BFF:bound} to establish a lower bound on Eve's knowledge about Alice's outcomes. Meanwhile, in the latter, the focus shifted towards choosing between the maximally entangled state and the CGLMP state as the nonlocal state utilized by Eve.

\begin{figure}[t]
\centering
\includegraphics[width=\columnwidth]{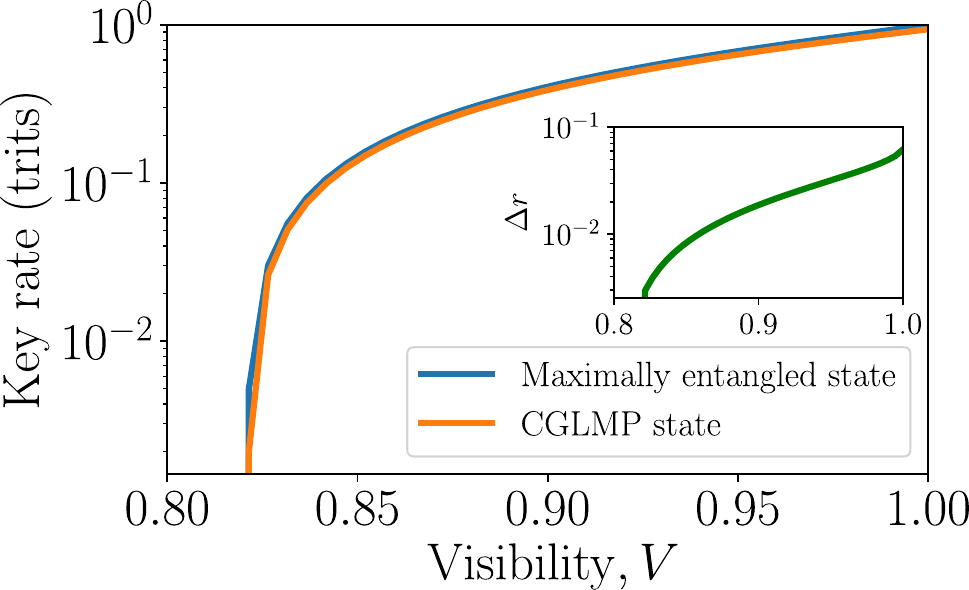}
\caption{CC-based upper bound on the key rate in terms of visibility when using the maximally entangled state and the CGLMP state for dimension $d=3$. In the inset plot, difference between both key rates is presented, i.e., $\Delta r = r_{\text{max}} - r_{\text{CGLMP}}$. For $V \gtrsim 0.805$, the upper bound is higher when using the maximally entangled state. The critical visibilities, for which $r_\textup{ub} = 0$, are $V_\textup{crit}^\textup{max} = 0.82043$ for the maximally entangled state, and $V_\textup{crit}^\textup{CGLMP} = 0.82101$ for the CGLMP state.}
\label{fig:r_vs_V}
\end{figure}

\begin{figure*}
    \centering
    \includegraphics[width=\textwidth]{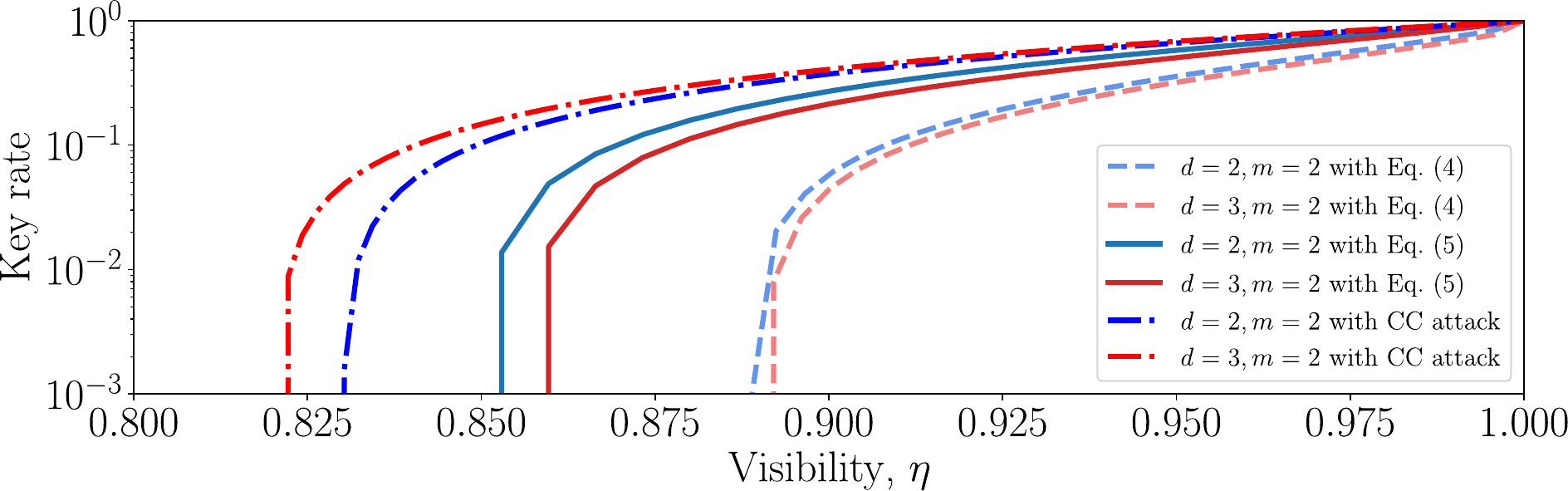}
    \caption{Comparison between the lower and upper bounds of the key rates for ($d=3,m=2$) depicted by the red curves, and ($d=2,m=2$) illustrated by the blue curves. The dash-dotted curves represent the upper bounds of the key rate, resulting in critical visibilities of $V^{(d=3)}_{\text{crit,ub}} \approx 0.820$ and $V^{(d=2)}_{\text{crit,ub}} = 0.830$. The lower bounds are displayed using solid and dashed curves. Specifically, the lower bounds obtained through Eq.~\eqref{Eq:min_entrpy} are represented by the dashed curves, yielding critical visibilities of $V^{(d=3)}_{\text{crit}} \approx 0.892$ and $V^{(d=2)}_{\text{crit}} \approx 0.888$. Conversely, the lower bounds derived from Eq.~\eqref{Eq:BFF:bound} are displayed as solid curves, resulting in critical visibilities of $V^{(d=3)}_{\text{crit,lb}} \approx 0.860$ and $V^{(d=2)}_{\text{crit,lb}} = 0.855$.}
    \label{Fig:all:rates:m2}
\end{figure*}

In relation to the lower bound, Fig.~\ref{Fig:lower:bounds} illustrates the impact of choosing between Eqs.~\eqref{Eq:min_entrpy} and \eqref{Eq:BFF:bound} on the behavior of the key rate concerning the visibility parameter $V$. Overall, across all studied cases, Eq.~\eqref{Eq:min_entrpy} consistently results in higher critical visibilities compared to Eq.~\eqref{Eq:BFF:bound}, which has been evaluated with $M=16$. Notably, significant observations emerge at this stage of analysis. Firstly, we note that the cases with $d=3$ (depicted by red curves) exhibit higher critical visibilities compared to those with $d=2$ (illustrated by blue curves), indicating that increasing the dimensions $d$ of the states utilized by Alice and Bob in the DIQKD protocol act in detriment of the key rate's noise robustness. Specifically, the critical visibilities obtained are approximately $V^{(d=2)}_{\text{crit}} \approx 0.888$ and $V^{(d=3)}_{\text{crit}} \approx 0.892$ when employing Eq.~\eqref{Eq:min_entrpy}, whereas $V^{(d=2)}_{\text{crit}} \approx 0.855$ and $V^{(d=3)}_{\text{crit}} \approx 0.860$ when using Eq.~\eqref{Eq:BFF:bound}. Secondly, although increasing the number of measurement settings does result in improvements in the obtained bounds, the enhancement achieved is nearly negligible. Consequently, we focus the rest of our analysis on the case of $m=2$.

On the other hand, Fig.~\ref{fig:r_vs_V} address the inquiry posed regarding the upper bound in the case $d=3$. Specifically, it illustrates the upper bound on the key rate computed considering both a maximally entangled state (blue curve) and a CGLMP state (orange curve). As can be observed, for the CGLMP state the upper bounds are slightly lower, and therefore the resilience to noise is marginally worse with respect to the maximally entangled state. This is further emphasized in the inset plot, depicting the difference between these rates as a function of visibility. Such a deviation in behaviour stems from an amplification of the EC-term when the CGLMP state is employed, since Alice and Bob's outcomes are less correlated than when the maximally entangled state is used. This leads to an overall decrease in the key rate and therefore a higher critical visibility (see SM~\ref{sec:additional_figures} for further details).

In Fig.~\ref{Fig:all:rates:m2}, a direct comparison between the upper (illustrated with dashed-dotted curves) and lower bounds (depicted with solid and dashed curves) is presented for the scenarios of $d=2$ and $d=3$, respectively denoted by blue and red colors. The most notable distinction in this plot is that, in contrast to the lower bounds, the critical visibilities derived from the upper bound are lower for both $d=3$ and $d=2$. Specifically, we obtain $V^{(d=3)}_{\text{crit,ub}} \approx 0.820$ and $V^{(d=2)}_{\text{crit,ub}} = 0.830$ for the upper bounds, while in the best case scenario lower bounds yield $V^{(d=3)}_{\text{crit,lb}} \approx 0.860$ and $V^{(d=2)}_{\text{crit,lb}} = 0.855$ using Eq.~\eqref{Eq:BFF:bound}. Additionally, a contrasting trend between the resulting bounds emerges from the preference for maximally entangled states in the analysis of lower bounds to the key rate, which optimally violate the Salavrakos' inequality, and the utilization of CGLMP states, which optimally violate the CGLMP-inequality. As detailed in SM~\ref{App:BFF}, the computation of the key rate for the lower bounds reveals two distinct regimes. For $V \gtrsim 0.901$, the Salavrakos' inequality yields superior lower bounds compared to the CGLMP inequality, which becomes dominant from this threshold until reaching the critical visibility. However, the optimization process depicted in Fig.~\ref{fig:scheme:optimization} slightly enhances these values, particularly for $V \lesssim 0.950$.

\begin{figure}[h!]
    \centering
    \includegraphics[width = 1\columnwidth]{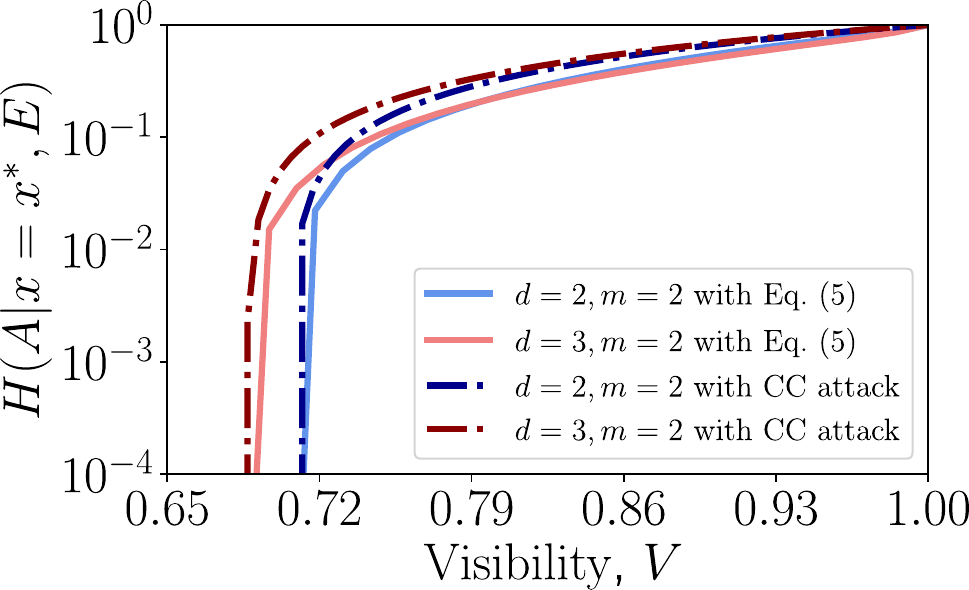}
    \caption{Comparison of lower bounds (solid curves) and upper bounds (dashdotted curves) for the EC-term $H(A|x=x^*,E)$ concerning $d=3$ (red curves) and $d=2$ (blue curves) as a function of visibility. The upper bounds are determined using the CGLMP state for the CC attack, yielding $V_{\text{crit,ub}}^{(d=3)} \approx 0.687$ and $V_{\text{crit,ub}}^{(d=2)} \approx 0.712$. Lower bounds are computed from Eq.~\eqref{Eq:BFF:bound}, resulting in $V_{\text{crit,lb}}^{(d=3)} \approx 0.691$ and $V_{\text{crit,lb}}^{(d=2)} \approx 0.713$.}
    \label{Fig:Entropy:comparison}
\end{figure}

Lastly, it is notable that the contrasting trend observed between lower and upper bounds in the key rate is not apparent when exclusively examining the PA-term. This is shown in Fig.~\ref{Fig:Entropy:comparison}, where the different upper and lower bounds to $H(A|x=x^*,E)$ presented throughout the text, are showcased as a function of the visibility. It is important to highlight that the upper bounds were evaluated utilizing the CGLMP state, due to its superior ability to bound $H(A|x=x^*,E)$ with respect to the visibility compared to maximally entangled states (see Appendix~\ref{sec:additional_figures}). As observed, both types of bounds exhibit enhanced critical visibility values for the $d=3$ case (red curves) in contrast to the $d=2$ case (blue curves). Specifically, for $d=3$ we find $V_{\text{crit,ub}}^{(d=3)} \approx 0.687$ and $V_{\text{crit,lb}}^{(d=3)} \approx 0.691$ , while for $d=2$ we obtain $V_{\text{crit,ub}}^{(d=2)} \approx 0.712$ and $V_{\text{crit,lb}}^{(d=2)} \approx 0.713$. Moreover, it is observed that the condition $H_{\text{ub}}^{(d=3)}(A|x=x^*,E) \geq H_{\text{ub}}^{(d=2)}(A|x=x^*,E)$ holds true across all visibilities for upper bounds but not for lower bounds. Particularly for the latter, $H_{\text{lb}}^{(d=3)}(A|x=x^*,E)$ and $ H_{\text{lb}}^{(d=2)}(A|x=x^*,E)$, the former assessed in terms of trits and the latter in bits, become equal at around $V \approx 0.795$. Nevertheless, this is a value for which the EC-term $H(A|B,x=x^*,y=y^*)$ already surpasses the PA-term, and therefore this enhancement is not reflected on the key rate. Thus, while an increase in $d$ does not assure a better key rate resilience concerning visibility, it does indeed hold promise for enhanced randomness generation.

\subsection{Upper bounds to the key rate for arbitrary dimensions}
\begin{figure}
\centering
\includegraphics[width=\columnwidth]{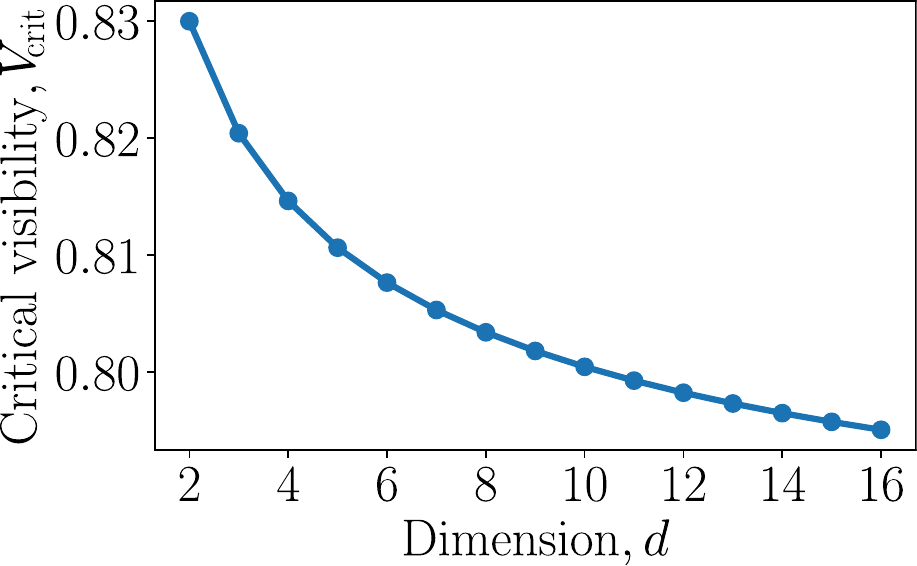}
\caption{Critical visibility $V_\textup{crit}$ obtained by means of Eq.~\eqref{Eq:r:up} of as a function of the dimension $d$. The critical visibility decreases rapidly at first and then increasingly slowly from $V_\textup{crit}\simeq0.8300$ for $d=2$ down to $V_\textup{crit} \simeq 0.7539$ for $d\to\infty$.}
\label{fig:Vcrit_vs_d}
\end{figure}

In contrast to the numerical analysis, where hardware limitations inevitably confine the dimension $d$ for computing a lower bound on the key rate, the analytical formulas derived for $r_{\text{ub}}$ when using the maximally entangled state allow us to transcend these constraints. Utilizing Eq.~\eqref{Eq:r:up}, we can determine a critical value of the visibility below which secure communication is unattainable for arbitrary dimensions $d \geq 2$ by setting $r_{\text{ub}} = 0$ and solving for $V$. This approach leads to the results presented in Fig.~\ref{fig:Vcrit_vs_d}, where the critical visibility is shown as a function of dimension $d$ ranging from $d=2$ to $d=16$. The plot illustrates a rapid decrease in this quantity initially, followed by a progressively slower decline as $d$ increases.

Furthermore, these derived bounds enable the examination of the critical visibility trend in the limit as $d$ approaches infinity. Specifically, by investigating this limit in Eq.~\eqref{Eq:r:up}, we deduce
\begin{equation}
    r_\textup{ub}^\infty = \lim_{d\to\infty} r_\textup{ub} = \frac{(2-\pi^2/(16\text{ Catalan}))V -1}{1-\pi^2/(16\text{ Catalan})}
\end{equation}
where we used that $\lim_{d\to\infty} I_d^\textup{max} = 32 \text{Catalan}/\pi^2\simeq 2.970$ \citep{collinsBellInequalitiesArbitrarily2002} ($\text{Catalan} \simeq 0.9159$ denotes Catalan's constant). By setting $r_\textup{ub}^\infty = 0$ and solving for $V$ we arrive at
\begin{equation}
    V_\textup{crit}^\infty = \frac{1}{2-\pi^2/(16\text{ Catalan})} \simeq 0.7539.
\end{equation}

This result allows us to obtain a critical value for the observed CGLMP-inequality violation below which no key exchange is possible using one-way communication reconciliation protocols. Specifically, given that $I_d^\textup{obs} = V I_d^\textup{max}$, we then find $\lim_{d\to\infty} I_d^\textup{crit} = V_\textup{crit}^\infty I_d^\textup{max} = 2.239$.

\section{CONCLUSIONS}
In this study, we examined security proofs of DIQKD protocols by contrasting the use of qubits and qutrits in states shared between Alice and Bob. Specifically, we optimized security proofs to establish lower bounds on the key rate in one-way communication DIQKD protocols. We compared these results with upper bounds on the key rate obtained by extending the CC attack to arbitrary dimensions. While the assessed lower bounds indicate that an increase in dimensionality does not yield improvements in the visibility requirements for achieving positive key rates, the opposite trend is observed for the upper bounds. Nevertheless, in both scenarios, an increase in dimensionality demonstrates an advantage concerning visibility constraints for device-independent randomness generation. Finally, the CC-attack-based proofs demonstrate an increase in noise tolerance concerning dimensionality, with enhancements plateauing at visibilities around 75\%. However, this limited improvement in tolerated visibility suggests that extending DIQKD experimental setups to higher dimensions beyond qubits might not warrant the associated increase in experimental complexity, at least in terms of noise robustness.

Future research avenues may explore the application of noisy preprocessing strategies~\cite{ho_noisy_2020} to ascertain potential improvements in the obtained bounds. Furthermore, it could motivate further analysis in the estimation of lower and upper bounds on the EC-terms.  Additionally, while our analysis has focused on the asymptotic limit, investigating finite-size scenarios could offer valuable insights into whether an increase in dimensionality proves beneficial or not.

\section*{ACKNOWLEDGEMENTS}
This work is supported by the Government of Spain (Severo Ochoa CEX2019-000910-S, FUNQIP, PRE2022-101475, NextGenerationEU PRTR-C17.I1), EU projects QSNP, Quantera Veriqtas and NEQST, Fundació Cellex, Fundació Mir-Puig, Generalitat de Catalunya (CERCA program), the ERC AdG CERQUTE and the AXA Chair in Quantum Information Science.

\bibliography{References.bib}{}

\begin{thebibliography}{42}%
\makeatletter
\providecommand \@ifxundefined [1]{%
 \@ifx{#1\undefined}
}%
\providecommand \@ifnum [1]{%
 \ifnum #1\expandafter \@firstoftwo
 \else \expandafter \@secondoftwo
 \fi
}%
\providecommand \@ifx [1]{%
 \ifx #1\expandafter \@firstoftwo
 \else \expandafter \@secondoftwo
 \fi
}%
\providecommand \natexlab [1]{#1}%
\providecommand \enquote  [1]{``#1''}%
\providecommand \bibnamefont  [1]{#1}%
\providecommand \bibfnamefont [1]{#1}%
\providecommand \citenamefont [1]{#1}%
\providecommand \href@noop [0]{\@secondoftwo}%
\providecommand \href [0]{\begingroup \@sanitize@url \@href}%
\providecommand \@href[1]{\@@startlink{#1}\@@href}%
\providecommand \@@href[1]{\endgroup#1\@@endlink}%
\providecommand \@sanitize@url [0]{\catcode `\\12\catcode `\$12\catcode `\&12\catcode `\#12\catcode `\^12\catcode `\_12\catcode `\%12\relax}%
\providecommand \@@startlink[1]{}%
\providecommand \@@endlink[0]{}%
\providecommand \url  [0]{\begingroup\@sanitize@url \@url }%
\providecommand \@url [1]{\endgroup\@href {#1}{\urlprefix }}%
\providecommand \urlprefix  [0]{URL }%
\providecommand \Eprint [0]{\href }%
\providecommand \doibase [0]{https://doi.org/}%
\providecommand \selectlanguage [0]{\@gobble}%
\providecommand \bibinfo  [0]{\@secondoftwo}%
\providecommand \bibfield  [0]{\@secondoftwo}%
\providecommand \translation [1]{[#1]}%
\providecommand \BibitemOpen [0]{}%
\providecommand \bibitemStop [0]{}%
\providecommand \bibitemNoStop [0]{.\EOS\space}%
\providecommand \EOS [0]{\spacefactor3000\relax}%
\providecommand \BibitemShut  [1]{\csname bibitem#1\endcsname}%
\let\auto@bib@innerbib\@empty
\bibitem [{\citenamefont {Aspect}(2023)}]{Benyoucef_book_ch2}%
  \BibitemOpen
  \bibfield  {author} {\bibinfo {author} {\bibfnamefont {A.}~\bibnamefont {Aspect}},\ }\bibfield  {title} {\bibinfo {title} {{The Second Quantum Revolution: From Basic Concepts to Quantum Technologies}},\ }in\ \href {https://doi.org/https://doi.org/10.1002/9783527837427.ch2} {\emph {\bibinfo {booktitle} {{Photonic Quantum Technologies}}}},\ \bibinfo {editor} {edited by\ \bibinfo {editor} {\bibfnamefont {M.}~\bibnamefont {Benyoucef}}}\ (\bibinfo  {publisher} {Wiley-VCH GmbH, Weinheim},\ \bibinfo {year} {2023})\ Chap.~\bibinfo {chapter} {2}, pp.\ \bibinfo {pages} {9--30}\BibitemShut {NoStop}%
\bibitem [{\citenamefont {Makarov}\ \emph {et~al.}(2006)\citenamefont {Makarov}, \citenamefont {Anisimov},\ and\ \citenamefont {Skaar}}]{makarov_effects_2006}%
  \BibitemOpen
  \bibfield  {author} {\bibinfo {author} {\bibfnamefont {V.}~\bibnamefont {Makarov}}, \bibinfo {author} {\bibfnamefont {A.}~\bibnamefont {Anisimov}},\ and\ \bibinfo {author} {\bibfnamefont {J.}~\bibnamefont {Skaar}},\ }\bibfield  {title} {\bibinfo {title} {Effects of detector efficiency mismatch on security of quantum cryptosystems},\ }\href {https://doi.org/10.1103/PhysRevA.74.022313} {\bibfield  {journal} {\bibinfo  {journal} {Physical Review A}\ }\textbf {\bibinfo {volume} {74}},\ \bibinfo {pages} {022313} (\bibinfo {year} {2006})},\ \Eprint {https://arxiv.org/abs/0511032} {arXiv:0511032} \BibitemShut {NoStop}%
\bibitem [{\citenamefont {Qi}\ \emph {et~al.}(2007)\citenamefont {Qi}, \citenamefont {Fung}, \citenamefont {Lo},\ and\ \citenamefont {Ma}}]{qi_time-shift_2007}%
  \BibitemOpen
  \bibfield  {author} {\bibinfo {author} {\bibfnamefont {B.}~\bibnamefont {Qi}}, \bibinfo {author} {\bibfnamefont {C.-H.~F.}\ \bibnamefont {Fung}}, \bibinfo {author} {\bibfnamefont {H.-K.}\ \bibnamefont {Lo}},\ and\ \bibinfo {author} {\bibfnamefont {X.}~\bibnamefont {Ma}},\ }\bibfield  {title} {\bibinfo {title} {Time-shift attack in practical quantum cryptosystems},\ }\href@noop {} {\bibfield  {journal} {\bibinfo  {journal} {Quantum Information \& Computation}\ }\textbf {\bibinfo {volume} {7}},\ \bibinfo {pages} {73} (\bibinfo {year} {2007})},\ \Eprint {https://arxiv.org/abs/0512080} {arXiv:0512080} \BibitemShut {NoStop}%
\bibitem [{\citenamefont {Zhao}\ \emph {et~al.}(2008)\citenamefont {Zhao}, \citenamefont {Fung}, \citenamefont {Qi}, \citenamefont {Chen},\ and\ \citenamefont {Lo}}]{zhao_quantum_2008}%
  \BibitemOpen
  \bibfield  {author} {\bibinfo {author} {\bibfnamefont {Y.}~\bibnamefont {Zhao}}, \bibinfo {author} {\bibfnamefont {C.-H.~F.}\ \bibnamefont {Fung}}, \bibinfo {author} {\bibfnamefont {B.}~\bibnamefont {Qi}}, \bibinfo {author} {\bibfnamefont {C.}~\bibnamefont {Chen}},\ and\ \bibinfo {author} {\bibfnamefont {H.-K.}\ \bibnamefont {Lo}},\ }\bibfield  {title} {\bibinfo {title} {Quantum hacking: {Experimental} demonstration of time-shift attack against practical quantum-key-distribution systems},\ }\href {https://doi.org/10.1103/PhysRevA.78.042333} {\bibfield  {journal} {\bibinfo  {journal} {Physical Review A}\ }\textbf {\bibinfo {volume} {78}},\ \bibinfo {pages} {042333} (\bibinfo {year} {2008})},\ \Eprint {https://arxiv.org/abs/0704.3253} {arXiv:0704.3253} \BibitemShut {NoStop}%
\bibitem [{\citenamefont {Makarov}(2009)}]{makarov_controlling_2009}%
  \BibitemOpen
  \bibfield  {author} {\bibinfo {author} {\bibfnamefont {V.}~\bibnamefont {Makarov}},\ }\bibfield  {title} {\bibinfo {title} {Controlling passively quenched single photon detectors by bright light},\ }\href {https://doi.org/10.1088/1367-2630/11/6/065003} {\bibfield  {journal} {\bibinfo  {journal} {New Journal of Physics}\ }\textbf {\bibinfo {volume} {11}},\ \bibinfo {pages} {065003} (\bibinfo {year} {2009})},\ \Eprint {https://arxiv.org/abs/0707.3987} {arXiv:0707.3987} \BibitemShut {NoStop}%
\bibitem [{\citenamefont {Lydersen}\ \emph {et~al.}(2010)\citenamefont {Lydersen}, \citenamefont {Wiechers}, \citenamefont {Wittmann}, \citenamefont {Elser}, \citenamefont {Skaar},\ and\ \citenamefont {Makarov}}]{lydersen_hacking_2010}%
  \BibitemOpen
  \bibfield  {author} {\bibinfo {author} {\bibfnamefont {L.}~\bibnamefont {Lydersen}}, \bibinfo {author} {\bibfnamefont {C.}~\bibnamefont {Wiechers}}, \bibinfo {author} {\bibfnamefont {C.}~\bibnamefont {Wittmann}}, \bibinfo {author} {\bibfnamefont {D.}~\bibnamefont {Elser}}, \bibinfo {author} {\bibfnamefont {J.}~\bibnamefont {Skaar}},\ and\ \bibinfo {author} {\bibfnamefont {V.}~\bibnamefont {Makarov}},\ }\bibfield  {title} {\bibinfo {title} {Hacking commercial quantum cryptography systems by tailored bright illumination},\ }\href {https://doi.org/10.1038/nphoton.2010.214} {\bibfield  {journal} {\bibinfo  {journal} {Nature Photonics}\ }\textbf {\bibinfo {volume} {4}},\ \bibinfo {pages} {686} (\bibinfo {year} {2010})},\ \Eprint {https://arxiv.org/abs/1008.4593} {arXiv:1008.4593} \BibitemShut {NoStop}%
\bibitem [{\citenamefont {Gerhardt}\ \emph {et~al.}(2011)\citenamefont {Gerhardt}, \citenamefont {Liu}, \citenamefont {Lamas-Linares}, \citenamefont {Skaar}, \citenamefont {Kurtsiefer},\ and\ \citenamefont {Makarov}}]{gerhardt_full-field_2011}%
  \BibitemOpen
  \bibfield  {author} {\bibinfo {author} {\bibfnamefont {I.}~\bibnamefont {Gerhardt}}, \bibinfo {author} {\bibfnamefont {Q.}~\bibnamefont {Liu}}, \bibinfo {author} {\bibfnamefont {A.}~\bibnamefont {Lamas-Linares}}, \bibinfo {author} {\bibfnamefont {J.}~\bibnamefont {Skaar}}, \bibinfo {author} {\bibfnamefont {C.}~\bibnamefont {Kurtsiefer}},\ and\ \bibinfo {author} {\bibfnamefont {V.}~\bibnamefont {Makarov}},\ }\bibfield  {title} {\bibinfo {title} {Full-field implementation of a perfect eavesdropper on a quantum cryptography system},\ }\href {https://doi.org/10.1038/ncomms1348} {\bibfield  {journal} {\bibinfo  {journal} {Nature Communications}\ }\textbf {\bibinfo {volume} {2}},\ \bibinfo {pages} {349} (\bibinfo {year} {2011})},\ \Eprint {https://arxiv.org/abs/1011.0105} {arXiv:1011.0105} \BibitemShut {NoStop}%
\bibitem [{\citenamefont {Ac\'{\i}n}\ \emph {et~al.}(2007)\citenamefont {Ac\'{\i}n}, \citenamefont {Brunner}, \citenamefont {Gisin}, \citenamefont {Massar}, \citenamefont {Pironio},\ and\ \citenamefont {Scarani}}]{Acin07}%
  \BibitemOpen
  \bibfield  {author} {\bibinfo {author} {\bibfnamefont {A.}~\bibnamefont {Ac\'{\i}n}}, \bibinfo {author} {\bibfnamefont {N.}~\bibnamefont {Brunner}}, \bibinfo {author} {\bibfnamefont {N.}~\bibnamefont {Gisin}}, \bibinfo {author} {\bibfnamefont {S.}~\bibnamefont {Massar}}, \bibinfo {author} {\bibfnamefont {S.}~\bibnamefont {Pironio}},\ and\ \bibinfo {author} {\bibfnamefont {V.}~\bibnamefont {Scarani}},\ }\bibfield  {title} {\bibinfo {title} {Device-independent security of quantum cryptography against collective attacks},\ }\href {https://doi.org/10.1103/PhysRevLett.98.230501} {\bibfield  {journal} {\bibinfo  {journal} {Phys. Rev. Lett.}\ }\textbf {\bibinfo {volume} {98}},\ \bibinfo {pages} {230501} (\bibinfo {year} {2007})}\BibitemShut {NoStop}%
\bibitem [{\citenamefont {Nadlinger}\ \emph {et~al.}(2022)\citenamefont {Nadlinger}, \citenamefont {Drmota}, \citenamefont {Nichol}, \citenamefont {Araneda}, \citenamefont {Main}, \citenamefont {Srinivas}, \citenamefont {Lucas}, \citenamefont {Ballance}, \citenamefont {Ivanov}, \citenamefont {Tan}, \citenamefont {Sekatski}, \citenamefont {Urbanke}, \citenamefont {Renner}, \citenamefont {Sangouard},\ and\ \citenamefont {Bancal}}]{nadlinger_experimental_2022}%
  \BibitemOpen
  \bibfield  {author} {\bibinfo {author} {\bibfnamefont {D.~P.}\ \bibnamefont {Nadlinger}}, \bibinfo {author} {\bibfnamefont {P.}~\bibnamefont {Drmota}}, \bibinfo {author} {\bibfnamefont {B.~C.}\ \bibnamefont {Nichol}}, \bibinfo {author} {\bibfnamefont {G.}~\bibnamefont {Araneda}}, \bibinfo {author} {\bibfnamefont {D.}~\bibnamefont {Main}}, \bibinfo {author} {\bibfnamefont {R.}~\bibnamefont {Srinivas}}, \bibinfo {author} {\bibfnamefont {D.~M.}\ \bibnamefont {Lucas}}, \bibinfo {author} {\bibfnamefont {C.~J.}\ \bibnamefont {Ballance}}, \bibinfo {author} {\bibfnamefont {K.}~\bibnamefont {Ivanov}}, \bibinfo {author} {\bibfnamefont {E.~Y.-Z.}\ \bibnamefont {Tan}}, \bibinfo {author} {\bibfnamefont {P.}~\bibnamefont {Sekatski}}, \bibinfo {author} {\bibfnamefont {R.~L.}\ \bibnamefont {Urbanke}}, \bibinfo {author} {\bibfnamefont {R.}~\bibnamefont {Renner}}, \bibinfo {author} {\bibfnamefont {N.}~\bibnamefont {Sangouard}},\ and\ \bibinfo {author} {\bibfnamefont {J.-D.}\ \bibnamefont {Bancal}},\ }\bibfield  {title}
  {\bibinfo {title} {Experimental quantum key distribution certified by {Bell}'s theorem},\ }\href {https://doi.org/10.1038/s41586-022-04941-5} {\bibfield  {journal} {\bibinfo  {journal} {Nature}\ }\textbf {\bibinfo {volume} {607}},\ \bibinfo {pages} {682} (\bibinfo {year} {2022})}\BibitemShut {NoStop}%
\bibitem [{\citenamefont {Zhang}\ \emph {et~al.}(2022)\citenamefont {Zhang}, \citenamefont {{van Leent}}, \citenamefont {Redeker}, \citenamefont {Garthoff}, \citenamefont {Schwonnek}, \citenamefont {Fertig}, \citenamefont {Eppelt}, \citenamefont {Scarani}, \citenamefont {Lim},\ and\ \citenamefont {Weinfurter}}]{zhangExperimentalDeviceindependentQuantum2022}%
  \BibitemOpen
  \bibfield  {author} {\bibinfo {author} {\bibfnamefont {W.}~\bibnamefont {Zhang}}, \bibinfo {author} {\bibfnamefont {T.}~\bibnamefont {{van Leent}}}, \bibinfo {author} {\bibfnamefont {K.}~\bibnamefont {Redeker}}, \bibinfo {author} {\bibfnamefont {R.}~\bibnamefont {Garthoff}}, \bibinfo {author} {\bibfnamefont {R.}~\bibnamefont {Schwonnek}}, \bibinfo {author} {\bibfnamefont {F.}~\bibnamefont {Fertig}}, \bibinfo {author} {\bibfnamefont {S.}~\bibnamefont {Eppelt}}, \bibinfo {author} {\bibfnamefont {V.}~\bibnamefont {Scarani}}, \bibinfo {author} {\bibfnamefont {C.~C.-W.}\ \bibnamefont {Lim}},\ and\ \bibinfo {author} {\bibfnamefont {H.}~\bibnamefont {Weinfurter}},\ }\bibfield  {title} {\bibinfo {title} {Experimental device-independent quantum key distribution between distant users},\ }\href {https://doi.org/10.1038/s41586-022-04891-y} {\bibfield  {journal} {\bibinfo  {journal} {Nature}\ }\textbf {\bibinfo {volume} {607}},\ \bibinfo {pages} {687} (\bibinfo {year} {2022})},\ \Eprint
  {https://arxiv.org/abs/2110.00575} {arxiv:2110.00575 [quant-ph]} \BibitemShut {NoStop}%
\bibitem [{\citenamefont {Liu}\ \emph {et~al.}(2022)\citenamefont {Liu}, \citenamefont {Zhang}, \citenamefont {Zhen}, \citenamefont {Li}, \citenamefont {Liu}, \citenamefont {Fan}, \citenamefont {Xu}, \citenamefont {Zhang},\ and\ \citenamefont {Pan}}]{China_exp_2022}%
  \BibitemOpen
  \bibfield  {author} {\bibinfo {author} {\bibfnamefont {W.-Z.}\ \bibnamefont {Liu}}, \bibinfo {author} {\bibfnamefont {Y.-Z.}\ \bibnamefont {Zhang}}, \bibinfo {author} {\bibfnamefont {Y.-Z.}\ \bibnamefont {Zhen}}, \bibinfo {author} {\bibfnamefont {M.-H.}\ \bibnamefont {Li}}, \bibinfo {author} {\bibfnamefont {Y.}~\bibnamefont {Liu}}, \bibinfo {author} {\bibfnamefont {J.}~\bibnamefont {Fan}}, \bibinfo {author} {\bibfnamefont {F.}~\bibnamefont {Xu}}, \bibinfo {author} {\bibfnamefont {Q.}~\bibnamefont {Zhang}},\ and\ \bibinfo {author} {\bibfnamefont {J.-W.}\ \bibnamefont {Pan}},\ }\bibfield  {title} {\bibinfo {title} {Toward a photonic demonstration of device-independent quantum key distribution},\ }\href {https://doi.org/10.1103/PhysRevLett.129.050502} {\bibfield  {journal} {\bibinfo  {journal} {Phys. Rev. Lett.}\ }\textbf {\bibinfo {volume} {129}},\ \bibinfo {pages} {050502} (\bibinfo {year} {2022})}\BibitemShut {NoStop}%
\bibitem [{\citenamefont {González-Ruiz}\ \emph {et~al.}(2022)\citenamefont {González-Ruiz}, \citenamefont {Rivera-Dean}, \citenamefont {Cenni}, \citenamefont {Sørensen}, \citenamefont {Acín},\ and\ \citenamefont {Oudot}}]{gonzalez-ruiz_device_2022}%
  \BibitemOpen
  \bibfield  {author} {\bibinfo {author} {\bibfnamefont {E.~M.}\ \bibnamefont {González-Ruiz}}, \bibinfo {author} {\bibfnamefont {J.}~\bibnamefont {Rivera-Dean}}, \bibinfo {author} {\bibfnamefont {M.~F.~B.}\ \bibnamefont {Cenni}}, \bibinfo {author} {\bibfnamefont {A.~S.}\ \bibnamefont {Sørensen}}, \bibinfo {author} {\bibfnamefont {A.}~\bibnamefont {Acín}},\ and\ \bibinfo {author} {\bibfnamefont {E.}~\bibnamefont {Oudot}},\ }\href {https://doi.org/10.48550/arXiv.2211.16472} {\bibinfo {title} {Device {Independent} {Quantum} {Key} {Distribution} with realistic single-photon source implementations}} (\bibinfo {year} {2022}),\ \bibinfo {note} {arXiv:2211.16472 [quant-ph]}\BibitemShut {NoStop}%
\bibitem [{\citenamefont {Arnon-Friedman}\ \emph {et~al.}(2019)\citenamefont {Arnon-Friedman}, \citenamefont {Renner},\ and\ \citenamefont {Vidick}}]{Friedman19}%
  \BibitemOpen
  \bibfield  {author} {\bibinfo {author} {\bibfnamefont {R.}~\bibnamefont {Arnon-Friedman}}, \bibinfo {author} {\bibfnamefont {R.}~\bibnamefont {Renner}},\ and\ \bibinfo {author} {\bibfnamefont {T.}~\bibnamefont {Vidick}},\ }\bibfield  {title} {\bibinfo {title} {Simple and tight device-independent security proofs},\ }\href {https://doi.org/10.1137/18M1174726} {\bibfield  {journal} {\bibinfo  {journal} {SIAM Journal on Computing}\ }\textbf {\bibinfo {volume} {48}},\ \bibinfo {pages} {181} (\bibinfo {year} {2019})},\ \Eprint {https://arxiv.org/abs/https://doi.org/10.1137/18M1174726} {https://doi.org/10.1137/18M1174726} \BibitemShut {NoStop}%
\bibitem [{\citenamefont {Clauser}\ \emph {et~al.}(1969)\citenamefont {Clauser}, \citenamefont {Horne}, \citenamefont {Shimony},\ and\ \citenamefont {Holt}}]{Clauser69}%
  \BibitemOpen
  \bibfield  {author} {\bibinfo {author} {\bibfnamefont {J.~F.}\ \bibnamefont {Clauser}}, \bibinfo {author} {\bibfnamefont {M.~A.}\ \bibnamefont {Horne}}, \bibinfo {author} {\bibfnamefont {A.}~\bibnamefont {Shimony}},\ and\ \bibinfo {author} {\bibfnamefont {R.~A.}\ \bibnamefont {Holt}},\ }\bibfield  {title} {\bibinfo {title} {Proposed experiment to test local hidden-variable theories},\ }\href {https://doi.org/10.1103/PhysRevLett.23.880} {\bibfield  {journal} {\bibinfo  {journal} {Phys. Rev. Lett.}\ }\textbf {\bibinfo {volume} {23}},\ \bibinfo {pages} {880} (\bibinfo {year} {1969})}\BibitemShut {NoStop}%
\bibitem [{\citenamefont {Mayers}\ and\ \citenamefont {Yao}(2004)}]{mayers_self_2004}%
  \BibitemOpen
  \bibfield  {author} {\bibinfo {author} {\bibfnamefont {D.}~\bibnamefont {Mayers}}\ and\ \bibinfo {author} {\bibfnamefont {A.}~\bibnamefont {Yao}},\ }\href {https://doi.org/10.48550/arXiv.quant-ph/0307205} {\bibinfo {title} {Self testing quantum apparatus}} (\bibinfo {year} {2004}),\ \bibinfo {note} {arXiv:quant-ph/0307205}\BibitemShut {NoStop}%
\bibitem [{\citenamefont {Kaniewski}(2016)}]{Kaniewski16}%
  \BibitemOpen
  \bibfield  {author} {\bibinfo {author} {\bibfnamefont {J.}~\bibnamefont {Kaniewski}},\ }\bibfield  {title} {\bibinfo {title} {Analytic and nearly optimal self-testing bounds for the clauser-horne-shimony-holt and mermin inequalities},\ }\href {https://doi.org/10.1103/PhysRevLett.117.070402} {\bibfield  {journal} {\bibinfo  {journal} {Phys. Rev. Lett.}\ }\textbf {\bibinfo {volume} {117}},\ \bibinfo {pages} {070402} (\bibinfo {year} {2016})}\BibitemShut {NoStop}%
\bibitem [{\citenamefont {Ho}\ \emph {et~al.}(2020)\citenamefont {Ho}, \citenamefont {Sekatski}, \citenamefont {Tan}, \citenamefont {Renner}, \citenamefont {Bancal},\ and\ \citenamefont {Sangouard}}]{ho_noisy_2020}%
  \BibitemOpen
  \bibfield  {author} {\bibinfo {author} {\bibfnamefont {M.}~\bibnamefont {Ho}}, \bibinfo {author} {\bibfnamefont {P.}~\bibnamefont {Sekatski}}, \bibinfo {author} {\bibfnamefont {E.-Z.}\ \bibnamefont {Tan}}, \bibinfo {author} {\bibfnamefont {R.}~\bibnamefont {Renner}}, \bibinfo {author} {\bibfnamefont {J.-D.}\ \bibnamefont {Bancal}},\ and\ \bibinfo {author} {\bibfnamefont {N.}~\bibnamefont {Sangouard}},\ }\bibfield  {title} {\bibinfo {title} {Noisy {Preprocessing} {Facilitates} a {Photonic} {Realization} of {Device}-{Independent} {Quantum} {Key} {Distribution}},\ }\href {https://doi.org/10.1103/PhysRevLett.124.230502} {\bibfield  {journal} {\bibinfo  {journal} {Physical Review Letters}\ }\textbf {\bibinfo {volume} {124}},\ \bibinfo {pages} {230502} (\bibinfo {year} {2020})},\ \Eprint {https://arxiv.org/abs/2005.13015} {arXiv:2005.13015} \BibitemShut {NoStop}%
\bibitem [{\citenamefont {Brown}\ \emph {et~al.}(2023)\citenamefont {Brown}, \citenamefont {Fawzi},\ and\ \citenamefont {Fawzi}}]{brown2023deviceindependent}%
  \BibitemOpen
  \bibfield  {author} {\bibinfo {author} {\bibfnamefont {P.}~\bibnamefont {Brown}}, \bibinfo {author} {\bibfnamefont {H.}~\bibnamefont {Fawzi}},\ and\ \bibinfo {author} {\bibfnamefont {O.}~\bibnamefont {Fawzi}},\ }\href@noop {} {\bibinfo {title} {Device-independent lower bounds on the conditional von neumann entropy}} (\bibinfo {year} {2023}),\ \Eprint {https://arxiv.org/abs/2106.13692} {arXiv:2106.13692 [quant-ph]} \BibitemShut {NoStop}%
\bibitem [{\citenamefont {Farkas}\ \emph {et~al.}(2021)\citenamefont {Farkas}, \citenamefont {Balanzó-Juandó}, \citenamefont {Łukanowski}, \citenamefont {Kołodyński},\ and\ \citenamefont {Acín}}]{farkas_bell_2021}%
  \BibitemOpen
  \bibfield  {author} {\bibinfo {author} {\bibfnamefont {M.}~\bibnamefont {Farkas}}, \bibinfo {author} {\bibfnamefont {M.}~\bibnamefont {Balanzó-Juandó}}, \bibinfo {author} {\bibfnamefont {K.}~\bibnamefont {Łukanowski}}, \bibinfo {author} {\bibfnamefont {J.}~\bibnamefont {Kołodyński}},\ and\ \bibinfo {author} {\bibfnamefont {A.}~\bibnamefont {Acín}},\ }\bibfield  {title} {\bibinfo {title} {Bell {Nonlocality} {Is} {Not} {Sufficient} for the {Security} of {Standard} {Device}-{Independent} {Quantum} {Key} {Distribution} {Protocols}},\ }\href {https://doi.org/10.1103/PhysRevLett.127.050503} {\bibfield  {journal} {\bibinfo  {journal} {Physical Review Letters}\ }\textbf {\bibinfo {volume} {127}},\ \bibinfo {pages} {050503} (\bibinfo {year} {2021})},\ \Eprint {https://arxiv.org/abs/2103.02639} {arXiv:2103.02639} \BibitemShut {NoStop}%
\bibitem [{\citenamefont {{\L}ukanowski}\ \emph {et~al.}(2023)\citenamefont {{\L}ukanowski}, \citenamefont {Farkas}, \citenamefont {{Balanz{\'o}-Juand{\'o}}}, \citenamefont {Ac{\'i}n},\ and\ \citenamefont {Ko{\l}ody{\'n}ski}}]{Lukanowski:23}%
  \BibitemOpen
  \bibfield  {author} {\bibinfo {author} {\bibfnamefont {K.}~\bibnamefont {{\L}ukanowski}}, \bibinfo {author} {\bibfnamefont {M.}~\bibnamefont {Farkas}}, \bibinfo {author} {\bibfnamefont {M.}~\bibnamefont {{Balanz{\'o}-Juand{\'o}}}}, \bibinfo {author} {\bibfnamefont {A.}~\bibnamefont {Ac{\'i}n}},\ and\ \bibinfo {author} {\bibfnamefont {J.}~\bibnamefont {Ko{\l}ody{\'n}ski}},\ }\bibfield  {title} {\bibinfo {title} {Upper bounds on key rates in device-independent quantum key distribution based on convex-combination attacks},\ }\href@noop {} {\bibfield  {journal} {\bibinfo  {journal} {Quantum}\ }\textbf {\bibinfo {volume} {7}},\ \bibinfo {pages} {1199} (\bibinfo {year} {2023})},\ \Eprint {https://arxiv.org/abs/2206.06245} {arxiv:2206.06245} \BibitemShut {NoStop}%
\bibitem [{\citenamefont {Sarkar}\ \emph {et~al.}(2021)\citenamefont {Sarkar}, \citenamefont {Saha}, \citenamefont {Andreas},\ and\ \citenamefont {Jędrzej-Augusiak}}]{Sarkar21}%
  \BibitemOpen
  \bibfield  {author} {\bibinfo {author} {\bibfnamefont {S.}~\bibnamefont {Sarkar}}, \bibinfo {author} {\bibfnamefont {D.}~\bibnamefont {Saha}}, \bibinfo {author} {\bibfnamefont {K.}~\bibnamefont {Andreas}},\ and\ \bibinfo {author} {\bibfnamefont {R.}~\bibnamefont {Jędrzej-Augusiak}},\ }\bibfield  {title} {\bibinfo {title} {Self-testing quantum systems of arbitrary local dimension with minimal number of measurements},\ }\href {https://doi.org/10.1038/s41534-021-00490-3} {\bibfield  {journal} {\bibinfo  {journal} {npj Quantum Information}\ }\textbf {\bibinfo {volume} {7}},\ \bibinfo {pages} {151} (\bibinfo {year} {2021})}\BibitemShut {NoStop}%
\bibitem [{\citenamefont {Collins}\ \emph {et~al.}(2002)\citenamefont {Collins}, \citenamefont {Gisin}, \citenamefont {Linden}, \citenamefont {Massar},\ and\ \citenamefont {Popescu}}]{collinsBellInequalitiesArbitrarily2002}%
  \BibitemOpen
  \bibfield  {author} {\bibinfo {author} {\bibfnamefont {D.}~\bibnamefont {Collins}}, \bibinfo {author} {\bibfnamefont {N.}~\bibnamefont {Gisin}}, \bibinfo {author} {\bibfnamefont {N.}~\bibnamefont {Linden}}, \bibinfo {author} {\bibfnamefont {S.}~\bibnamefont {Massar}},\ and\ \bibinfo {author} {\bibfnamefont {S.}~\bibnamefont {Popescu}},\ }\bibfield  {title} {\bibinfo {title} {Bell inequalities for arbitrarily high dimensional systems},\ }\href {https://doi.org/10.1103/PhysRevLett.88.040404} {\bibfield  {journal} {\bibinfo  {journal} {Physical Review Letters}\ }\textbf {\bibinfo {volume} {88}},\ \bibinfo {pages} {040404} (\bibinfo {year} {2002})},\ \Eprint {https://arxiv.org/abs/quant-ph/0106024} {arxiv:quant-ph/0106024} \BibitemShut {NoStop}%
\bibitem [{\citenamefont {Salavrakos}\ \emph {et~al.}(2017)\citenamefont {Salavrakos}, \citenamefont {Augusiak}, \citenamefont {Tura}, \citenamefont {Wittek}, \citenamefont {Ac{\'i}n},\ and\ \citenamefont {Pironio}}]{salavrakosBellInequalitiesTailored2017}%
  \BibitemOpen
  \bibfield  {author} {\bibinfo {author} {\bibfnamefont {A.}~\bibnamefont {Salavrakos}}, \bibinfo {author} {\bibfnamefont {R.}~\bibnamefont {Augusiak}}, \bibinfo {author} {\bibfnamefont {J.}~\bibnamefont {Tura}}, \bibinfo {author} {\bibfnamefont {P.}~\bibnamefont {Wittek}}, \bibinfo {author} {\bibfnamefont {A.}~\bibnamefont {Ac{\'i}n}},\ and\ \bibinfo {author} {\bibfnamefont {S.}~\bibnamefont {Pironio}},\ }\bibfield  {title} {\bibinfo {title} {Bell inequalities tailored to maximally entangled states},\ }\href {https://doi.org/10.1103/PhysRevLett.119.040402} {\bibfield  {journal} {\bibinfo  {journal} {Physical Review Letters}\ }\textbf {\bibinfo {volume} {119}},\ \bibinfo {pages} {040402} (\bibinfo {year} {2017})},\ \Eprint {https://arxiv.org/abs/1607.04578} {arxiv:1607.04578 [quant-ph]} \BibitemShut {NoStop}%
\bibitem [{\citenamefont {Cerf}\ \emph {et~al.}(2002)\citenamefont {Cerf}, \citenamefont {Bourennane}, \citenamefont {Karlsson},\ and\ \citenamefont {Gisin}}]{Cerf_qudit_QKD_2002}%
  \BibitemOpen
  \bibfield  {author} {\bibinfo {author} {\bibfnamefont {N.~J.}\ \bibnamefont {Cerf}}, \bibinfo {author} {\bibfnamefont {M.}~\bibnamefont {Bourennane}}, \bibinfo {author} {\bibfnamefont {A.}~\bibnamefont {Karlsson}},\ and\ \bibinfo {author} {\bibfnamefont {N.}~\bibnamefont {Gisin}},\ }\bibfield  {title} {\bibinfo {title} {Security of quantum key distribution using $\mathit{d}$-level systems},\ }\href {https://doi.org/10.1103/PhysRevLett.88.127902} {\bibfield  {journal} {\bibinfo  {journal} {Phys. Rev. Lett.}\ }\textbf {\bibinfo {volume} {88}},\ \bibinfo {pages} {127902} (\bibinfo {year} {2002})}\BibitemShut {NoStop}%
\bibitem [{\citenamefont {Borkała}\ \emph {et~al.}(2022)\citenamefont {Borkała}, \citenamefont {Jebarathinam}, \citenamefont {Sarkar},\ and\ \citenamefont {Augusiak}}]{randomness_qutrit}%
  \BibitemOpen
  \bibfield  {author} {\bibinfo {author} {\bibfnamefont {J.~J.}\ \bibnamefont {Borkała}}, \bibinfo {author} {\bibfnamefont {C.}~\bibnamefont {Jebarathinam}}, \bibinfo {author} {\bibfnamefont {S.}~\bibnamefont {Sarkar}},\ and\ \bibinfo {author} {\bibfnamefont {R.}~\bibnamefont {Augusiak}},\ }\bibfield  {title} {\bibinfo {title} {Device-independent certification of maximal randomness from pure entangled two-qutrit states using non-projective measurements},\ }\bibfield  {journal} {\bibinfo  {journal} {Entropy}\ }\textbf {\bibinfo {volume} {24}},\ \href {https://doi.org/10.3390/e24030350} {10.3390/e24030350} (\bibinfo {year} {2022})\BibitemShut {NoStop}%
\bibitem [{\citenamefont {Renner}\ \emph {et~al.}(2005)\citenamefont {Renner}, \citenamefont {Gisin},\ and\ \citenamefont {Kraus}}]{renner_information-theoretic_2005}%
  \BibitemOpen
  \bibfield  {author} {\bibinfo {author} {\bibfnamefont {R.}~\bibnamefont {Renner}}, \bibinfo {author} {\bibfnamefont {N.}~\bibnamefont {Gisin}},\ and\ \bibinfo {author} {\bibfnamefont {B.}~\bibnamefont {Kraus}},\ }\bibfield  {title} {\bibinfo {title} {Information-theoretic security proof for quantum-key-distribution protocols},\ }\href {https://doi.org/10.1103/PhysRevA.72.012332} {\bibfield  {journal} {\bibinfo  {journal} {Physical Review A}\ }\textbf {\bibinfo {volume} {72}},\ \bibinfo {pages} {012332} (\bibinfo {year} {2005})},\ \Eprint {https://arxiv.org/abs/0502064} {arXiv:0502064} \BibitemShut {NoStop}%
\bibitem [{\citenamefont {Navascués}\ \emph {et~al.}(2007)\citenamefont {Navascués}, \citenamefont {Pironio},\ and\ \citenamefont {Acín}}]{navascues_bounding_2007}%
  \BibitemOpen
  \bibfield  {author} {\bibinfo {author} {\bibfnamefont {M.}~\bibnamefont {Navascués}}, \bibinfo {author} {\bibfnamefont {S.}~\bibnamefont {Pironio}},\ and\ \bibinfo {author} {\bibfnamefont {A.}~\bibnamefont {Acín}},\ }\bibfield  {title} {\bibinfo {title} {Bounding the {Set} of {Quantum} {Correlations}},\ }\href {https://doi.org/10.1103/PhysRevLett.98.010401} {\bibfield  {journal} {\bibinfo  {journal} {Physical Review Letters}\ }\textbf {\bibinfo {volume} {98}},\ \bibinfo {pages} {010401} (\bibinfo {year} {2007})},\ \Eprint {https://arxiv.org/abs/0607119} {arXiv:0607119} \BibitemShut {NoStop}%
\bibitem [{\citenamefont {Navascués}\ \emph {et~al.}(2008)\citenamefont {Navascués}, \citenamefont {Pironio},\ and\ \citenamefont {Acín}}]{navascues_convergent_2008}%
  \BibitemOpen
  \bibfield  {author} {\bibinfo {author} {\bibfnamefont {M.}~\bibnamefont {Navascués}}, \bibinfo {author} {\bibfnamefont {S.}~\bibnamefont {Pironio}},\ and\ \bibinfo {author} {\bibfnamefont {A.}~\bibnamefont {Acín}},\ }\bibfield  {title} {\bibinfo {title} {A convergent hierarchy of semidefinite programs characterizing the set of quantum correlations},\ }\href {https://doi.org/10.1088/1367-2630/10/7/073013} {\bibfield  {journal} {\bibinfo  {journal} {New Journal of Physics}\ }\textbf {\bibinfo {volume} {10}},\ \bibinfo {pages} {073013} (\bibinfo {year} {2008})},\ \Eprint {https://arxiv.org/abs/0803.4290} {arXiv:0803.4290} \BibitemShut {NoStop}%
\bibitem [{\citenamefont {Brunner}\ \emph {et~al.}(2014)\citenamefont {Brunner}, \citenamefont {Cavalcanti}, \citenamefont {Pironio}, \citenamefont {Scarani},\ and\ \citenamefont {Wehner}}]{brunnerBellNonlocality2014}%
  \BibitemOpen
  \bibfield  {author} {\bibinfo {author} {\bibfnamefont {N.}~\bibnamefont {Brunner}}, \bibinfo {author} {\bibfnamefont {D.}~\bibnamefont {Cavalcanti}}, \bibinfo {author} {\bibfnamefont {S.}~\bibnamefont {Pironio}}, \bibinfo {author} {\bibfnamefont {V.}~\bibnamefont {Scarani}},\ and\ \bibinfo {author} {\bibfnamefont {S.}~\bibnamefont {Wehner}},\ }\bibfield  {title} {\bibinfo {title} {Bell nonlocality},\ }\href {https://doi.org/10.1103/RevModPhys.86.419} {\bibfield  {journal} {\bibinfo  {journal} {Reviews of Modern Physics}\ }\textbf {\bibinfo {volume} {86}},\ \bibinfo {pages} {419} (\bibinfo {year} {2014})},\ \Eprint {https://arxiv.org/abs/1303.2849} {arxiv:1303.2849 [quant-ph]} \BibitemShut {NoStop}%
\bibitem [{\citenamefont {Devetak}\ and\ \citenamefont {Winter}(2005)}]{Devetak_2005}%
  \BibitemOpen
  \bibfield  {author} {\bibinfo {author} {\bibfnamefont {I.}~\bibnamefont {Devetak}}\ and\ \bibinfo {author} {\bibfnamefont {A.}~\bibnamefont {Winter}},\ }\bibfield  {title} {\bibinfo {title} {Distillation of secret key and entanglement from quantum states},\ }\href {https://doi.org/10.1098/rspa.2004.1372} {\bibfield  {journal} {\bibinfo  {journal} {Proceedings of the Royal Society A: Mathematical, Physical and Engineering Sciences}\ }\textbf {\bibinfo {volume} {461}},\ \bibinfo {pages} {207–235} (\bibinfo {year} {2005})}\BibitemShut {NoStop}%
\bibitem [{\citenamefont {Konig}\ \emph {et~al.}(2009)\citenamefont {Konig}, \citenamefont {Renner},\ and\ \citenamefont {Schaffner}}]{konig_operational_2009}%
  \BibitemOpen
  \bibfield  {author} {\bibinfo {author} {\bibfnamefont {R.}~\bibnamefont {Konig}}, \bibinfo {author} {\bibfnamefont {R.}~\bibnamefont {Renner}},\ and\ \bibinfo {author} {\bibfnamefont {C.}~\bibnamefont {Schaffner}},\ }\bibfield  {title} {\bibinfo {title} {The {Operational} {Meaning} of {Min}- and {Max}-{Entropy}},\ }\href {https://doi.org/10.1109/TIT.2009.2025545} {\bibfield  {journal} {\bibinfo  {journal} {IEEE Transactions on Information Theory}\ }\textbf {\bibinfo {volume} {55}},\ \bibinfo {pages} {4337} (\bibinfo {year} {2009})},\ \Eprint {https://arxiv.org/abs/0807.1338} {arXiv:0807.1338} \BibitemShut {NoStop}%
\bibitem [{\citenamefont {Acín}\ \emph {et~al.}(2012)\citenamefont {Acín}, \citenamefont {Massar},\ and\ \citenamefont {Pironio}}]{acin_randomness_2012}%
  \BibitemOpen
  \bibfield  {author} {\bibinfo {author} {\bibfnamefont {A.}~\bibnamefont {Acín}}, \bibinfo {author} {\bibfnamefont {S.}~\bibnamefont {Massar}},\ and\ \bibinfo {author} {\bibfnamefont {S.}~\bibnamefont {Pironio}},\ }\bibfield  {title} {\bibinfo {title} {Randomness versus {Nonlocality} and {Entanglement}},\ }\href {https://doi.org/10.1103/PhysRevLett.108.100402} {\bibfield  {journal} {\bibinfo  {journal} {Physical Review Letters}\ }\textbf {\bibinfo {volume} {108}},\ \bibinfo {pages} {100402} (\bibinfo {year} {2012})},\ \Eprint {https://arxiv.org/abs/1107.2754} {arXiv:1107.2754} \BibitemShut {NoStop}%
\bibitem [{\citenamefont {Boyd}\ and\ \citenamefont {Lieven}(2004)}]{BoydCH4}%
  \BibitemOpen
  \bibfield  {author} {\bibinfo {author} {\bibfnamefont {S.}~\bibnamefont {Boyd}}\ and\ \bibinfo {author} {\bibfnamefont {V.}~\bibnamefont {Lieven}},\ }\bibinfo {title} {Convex optimization problems},\ in\ \href@noop {} {\emph {\bibinfo {booktitle} {{Convex Optimization}}}}\ (\bibinfo  {publisher} {Cambridge University Press, Cambridge, UK},\ \bibinfo {year} {2004})\ Chap.~\bibinfo {chapter} {4}, pp.\ \bibinfo {pages} {127--214}\BibitemShut {NoStop}%
\bibitem [{\citenamefont {Acin}\ \emph {et~al.}(2002)\citenamefont {Acin}, \citenamefont {Durt}, \citenamefont {Gisin},\ and\ \citenamefont {Latorre}}]{acinQuantumNonlocalityTwo2002}%
  \BibitemOpen
  \bibfield  {author} {\bibinfo {author} {\bibfnamefont {A.}~\bibnamefont {Acin}}, \bibinfo {author} {\bibfnamefont {T.}~\bibnamefont {Durt}}, \bibinfo {author} {\bibfnamefont {N.}~\bibnamefont {Gisin}},\ and\ \bibinfo {author} {\bibfnamefont {J.~I.}\ \bibnamefont {Latorre}},\ }\bibfield  {title} {\bibinfo {title} {Quantum non-locality in two three-level systems},\ }\href {https://doi.org/10.1103/PhysRevA.65.052325} {\bibfield  {journal} {\bibinfo  {journal} {Physical Review A}\ }\textbf {\bibinfo {volume} {65}},\ \bibinfo {pages} {052325} (\bibinfo {year} {2002})},\ \Eprint {https://arxiv.org/abs/quant-ph/0111143} {arxiv:quant-ph/0111143} \BibitemShut {NoStop}%
\bibitem [{\citenamefont {Masanes}(2003)}]{DBLP:journals/qic/Masanes03}%
  \BibitemOpen
  \bibfield  {author} {\bibinfo {author} {\bibfnamefont {L.}~\bibnamefont {Masanes}},\ }\bibfield  {title} {\bibinfo {title} {Tight {{Bell}} inequality for d-outcome measurements correlations},\ }\href {https://doi.org/10.26421/QIC3.4-4} {\bibfield  {journal} {\bibinfo  {journal} {Quantum Information \& Computation}\ }\textbf {\bibinfo {volume} {3}},\ \bibinfo {pages} {345} (\bibinfo {year} {2003})}\BibitemShut {NoStop}%
\bibitem [{\citenamefont {Barrett}\ \emph {et~al.}(2006)\citenamefont {Barrett}, \citenamefont {Kent},\ and\ \citenamefont {Pironio}}]{barrett_maximally_2006}%
  \BibitemOpen
  \bibfield  {author} {\bibinfo {author} {\bibfnamefont {J.}~\bibnamefont {Barrett}}, \bibinfo {author} {\bibfnamefont {A.}~\bibnamefont {Kent}},\ and\ \bibinfo {author} {\bibfnamefont {S.}~\bibnamefont {Pironio}},\ }\bibfield  {title} {\bibinfo {title} {Maximally {Nonlocal} and {Monogamous} {Quantum} {Correlations}},\ }\href {https://doi.org/10.1103/PhysRevLett.97.170409} {\bibfield  {journal} {\bibinfo  {journal} {Physical Review Letters}\ }\textbf {\bibinfo {volume} {97}},\ \bibinfo {pages} {170409} (\bibinfo {year} {2006})},\ \Eprint {https://arxiv.org/abs/0605182} {arXiv:0605182} \BibitemShut {NoStop}%
\bibitem [{\citenamefont {Lofberg}(2004)}]{lofberg_yalmip_2004}%
  \BibitemOpen
  \bibfield  {author} {\bibinfo {author} {\bibfnamefont {J.}~\bibnamefont {Lofberg}},\ }\bibfield  {title} {\bibinfo {title} {{YALMIP} : a toolbox for modeling and optimization in {MATLAB}},\ }in\ \href {https://doi.org/10.1109/CACSD.2004.1393890} {\emph {\bibinfo {booktitle} {2004 {IEEE} {International} {Conference} on {Robotics} and {Automation} ({IEEE} {Cat}. {No}.{04CH37508})}}}\ (\bibinfo {year} {2004})\ pp.\ \bibinfo {pages} {284--289}\BibitemShut {NoStop}%
\bibitem [{\citenamefont {ApS}(2019)}]{mosek}%
  \BibitemOpen
  \bibfield  {author} {\bibinfo {author} {\bibfnamefont {M.}~\bibnamefont {ApS}},\ }\href {http://docs.mosek.com/9.0/toolbox/index.html} {\emph {\bibinfo {title} {The MOSEK optimization toolbox for MATLAB manual. Version 9.0.}}} (\bibinfo {year} {2019})\BibitemShut {NoStop}%
\bibitem [{\citenamefont {{The MathWorks Inc.}}(2022)}]{MatlabGO}%
  \BibitemOpen
  \bibfield  {author} {\bibinfo {author} {\bibnamefont {{The MathWorks Inc.}}},\ }\href {https://es.mathworks.com/help/gads/ref/globalOptimization.html} {\bibinfo {title} {{Global Optimization Toolbox (R2022a)}}} (\bibinfo {year} {2022})\BibitemShut {NoStop}%
\bibitem [{\citenamefont {Wittek}(2015)}]{wittek_algorithm_2015}%
  \BibitemOpen
  \bibfield  {author} {\bibinfo {author} {\bibfnamefont {P.}~\bibnamefont {Wittek}},\ }\bibfield  {title} {\bibinfo {title} {Algorithm 950: {Ncpol2sdpa} --{Sparse} {Semidefinite} {Programming} {Relaxations} for {Polynomial} {Optimization} {Problems} of {Noncommuting} {Variables}},\ }\href {https://doi.org/10.1145/2699464} {\bibfield  {journal} {\bibinfo  {journal} {ACM Transactions on Mathematical Software}\ }\textbf {\bibinfo {volume} {41}},\ \bibinfo {pages} {21:1} (\bibinfo {year} {2015})}\BibitemShut {NoStop}%
\bibitem [{\citenamefont {Brown}(2021)}]{PBrown_Github}%
  \BibitemOpen
  \bibfield  {author} {\bibinfo {author} {\bibfnamefont {P.}~\bibnamefont {Brown}},\ }\href@noop {} {\bibinfo {title} {Example scripts for computing rates of device-independent protocols}},\ \bibinfo {howpublished} {\url{https://github.com/peterjbrown519/DI-rates}} (\bibinfo {year} {2021})\BibitemShut {NoStop}%
\bibitem [{\citenamefont {{ApS}}(2022)}]{mosekpython}%
  \BibitemOpen
  \bibfield  {author} {\bibinfo {author} {\bibfnamefont {M.}~\bibnamefont {{ApS}}},\ }\href@noop {} {\bibinfo {title} {{MOSEK} {Optimizer} {API} for {Python} 9.3.20}},\ \bibinfo {howpublished} {\url{https://docs.mosek.com/latest/pythonapi/index.html}} (\bibinfo {year} {2022})\BibitemShut {NoStop}%
\end{thebibliography}%

\clearpage
\onecolumngrid
\appendix

\begin{center}
\large{\textbf{\textsc{Supplementary Material}}}
\end{center}
\section{Parameterization of the employed measurements}\label{App:measurements}
When working with $d=2$, the set of projective operators we considered for characterizing Alice's measurements (same for Bob) is spanned by
\begin{equation}
    \big\{\hat{\Pi}_{0|x}=\dyad{\psi_{2}},
    \hat{\Pi}_{1|x}=\mathbb{1}-\hat{\Pi}_{0|x}
    \big\}
\end{equation}
where we parameterize the state $\ket{\psi_2}$ in the equation above as
\begin{equation}
\ket{\psi_{2}}=\cos\theta^{(x)}\ket{0}+e^{i\phi^{(x)}}\sin\theta_x\ket{1},
\end{equation}
which depends on two parameters $(\theta^{(x)}, \phi^{(x)})$. Thus, the total number of parameters used in the case Alice and Bob respectively implement $m$ and $m+1$ measurements, is $2 m (m+1)$.

As for $d=3$, the set of projective operators spanning the measurements applied by both parties is given by
\begin{equation}
    \big\{\hat{\Pi}_{0|x}=\dyad{\psi_{3}},
    \{\hat{\Pi}_{1|x}=\dyad{\psi^{\perp}_{3}},
    \hat{\Pi}_{2|x}=\mathbb{1}-\hat{\Pi}_{0|x} - \hat{\Pi}_{1|x}
    \big\},
\end{equation}
where, in this case, we express the state $\ket{\psi_3}$ as
\begin{equation}
    \ket{\psi_{3}} = \cos\phi_0 \sin\theta_0\ket{0} + 
    e^{-\frac{2}{3} i \alpha_0 \pi} \sin\phi_0 \sin\theta_0\ket{1} +
    e^{-\frac{4}{3} i \beta_0 \pi} \cos\theta_0\ket{2},
\end{equation}
where we have omitted the the index $x$ from the settings for simplicity. In any case, this parameterization must be done for each value of the input $x$. Then, by means of the the Gram–Schmidt process we can find a parameterized state orthonormal to $\ket{\psi_{3}}$, which reads as
\begin{equation}
    \begin{aligned}
    \ket{\psi_3^\perp}
    & = \frac{1}{\mathcal{N}} \bigg[
    \bigg( \cos\phi_1 \sin\theta_1 \left( \sin^2\phi_0 \sin^2\theta_0 +\cos^2\theta_0 \right) \\
    &\quad - \cos\phi_0 \sin\theta_0 \left(e^{\frac{2}{3} \pi i  (\alpha_0 - \alpha_1 + 1)} \sin\phi_0 \sin\phi_1 \sin\theta_0 \sin\theta_1 + e^{\frac{4}{3} \pi i  (\beta_0 - \beta_1 + 1)} \cos\theta_0 \cos\theta_1\right) \bigg) \ket{0}\\
    &\quad +  e^{-\frac{2}{3}  \pi i  \alpha_0} \Big( \cos\phi_0 \sin^2\theta_0 \sin\theta_1 \left(-\sin\phi_0 \cos\phi_1 + (-1)^{2/3} e^{\frac{2}{3} \pi i  (\alpha_0-\alpha_1)} \cos\phi_0 \sin\phi_1 \right)\\
    &\quad + (-1)^{2/3} e^{\frac{2}{3} \pi i (\alpha_0 - \alpha_1)} \sin\phi_1 \cos^2\theta_0 \sin\theta_1 + (-1)^{1/3} e^{\frac{4}{3} \pi i  (\beta_0-\beta_1)} \sin\phi_0 \sin\theta_0 \cos\theta_0 \cos\theta_1 \Big)  \ket{1}\\
    &\quad + e^{-\frac{4}{3} \pi i \beta_0} \sin\theta_0 \bigg( \cos\theta_0 \sin\theta_1 \left( \cos\phi_0 \cos\phi_1+(-1)^{2/3} e^{\frac{2}{3} \pi i  (\alpha_0-\alpha_1)} \sin\phi_0 \sin\phi_1 \right)\\
    &\quad +(-1)^{1/3} e^{\frac{4}{3} \pi i  (\beta_0-\beta_1)} \sin\theta_0 \cos\theta_1 \bigg) \ket{2}
    \bigg],
    \end{aligned}
\end{equation}
where $\mathcal{N}$ is the normalization. 

Thus, this measurement depends on a total of 8 parameters, implying that the total number of parameters given that Alice and Bob respectively apply $m$ and $m+1$ measurements is $8m(m+1)$.

\section{Optimizing the key rate through $H_{\text{min}}(A|E)$}\label{App:Optimization}

In this section, we provide a step-by-step description of the methodology used for optimizing the lower bound on the key rate of the DIQKD protocol outlined in Fig.~\ref{fig:scheme:optimization}. This optimization essentially consists of an optimization over the variables defining Alice and Bob's measurements. Hereupon, we denote the corresponding POVM sets as $M_{x}(\boldsymbol{\theta}_x)\coloneqq\{\hat{\Pi}_{a\vert x}(\boldsymbol{\theta}_x)\}_a$ and $M_{x}(\boldsymbol{\theta}_y)\coloneqq\{\hat{\Pi}_{b\vert y}(\boldsymbol{\theta}_y)\}_b$, where $\hat{\Pi}_{a\vert x}(\boldsymbol{\theta}_x)$ are and $\hat{\Pi}_{b\vert y}(\boldsymbol{\theta}_y)$ are projective operators for Alice and Bob respectively, while $\boldsymbol{\theta}_x$ and $\boldsymbol{\theta}_y$ correspond to the employed parameters (for more details about the parameterization of the measurements see SM~\ref{App:measurements}). The length of these vectors depends on the dimensions of Alice and Bob's Hilbert spaces. As mentioned in the previous subsection, for the case of qubits each vector has two elements, while for qutrits they have eight elements. 

The optimization method followed here comprises several steps. These alternate between SDP optimizations defining the min-entropy as in Eq.~\eqref{Eq:min_entrpy}, and a local-optimization-based approach on the $\boldsymbol{\theta}_x$ and $\boldsymbol{\theta}_y$ vectors. More specifically, these steps are:
\begin{enumerate}
    \item We begin by fixing the amount of noise we allow on the state, that is the visibility $V$, leading to
        \begin{equation}
        \hat{\rho}_0
            = 
                    V
                    \dyad{\psi_0}
                    + \dfrac{(1-V)}{d} \mathbbm{1},
    \end{equation}
    as well as the measurement settings $\boldsymbol{\theta}^{(0)}_x$ and $\boldsymbol{\theta}^{(0)}_y$, and the initial state $\ket{\psi_0}$. For the zeroth step of the optimization technique, we consider the ideal scenario $V = 1.0$ and set $\ket{\psi_0} = (1/\sqrt{d})\sum_{q=1}^{d
     }\ket{qq}$, i.e. the maximally entangled state, for which we know that Alice and Bob can optimally maximize the key rate~\cite{salavrakosBellInequalitiesTailored2017}, using measurement settings coinciding with those maximizing the Bell inequalities presented in Refs.~\cite{salavrakosBellInequalitiesTailored2017,barrett_maximally_2006}.

    These parameters are updated after each iteration of the algorithm, until reaching the minimum value of $V = 0.8$, for which the optimal value of the key rate already becomes negative.
    
    \item The previous step of the algorithm allows us to compute the conditional probabilities $\{p(a,b|x,y)\}$. These are then used as constraints for the convex-optimization problem 
    \begin{equation}
        \begin{aligned}
        G(A|x=x^*,E)=\underset{Z_a}{\text{sup}} &\sum_a \text{Tr}[ \hat{\rho}_{ABE}(\hat{\Pi}_{a|x^*} Z_a)\big]\\
        \text{s.t.} &\; \text{Tr} \big[ \hat{\rho}_{ABE}(\hat{\Pi}_{a|x}\hat{\Pi}_{b|y} )\big]=p(a,b|x,y)\\
        &\sum_a \hat{\Pi}_{a|x}=\sum_b \hat{\Pi}_{b|y}=\mathbb{1} &&\forall\; x,y\\
        & \;\hat{\Pi}_{a|x}\geq 0, \quad \hat{\Pi}_{b|y}\geq 0 &&\forall\; a,b,x,y\\
        & \;\hat{\Pi}^2_{a|x}=\hat{\Pi}_{a|x} &&\forall\; a,x\\
        & \;\hat{\Pi}^2_{b|y}=\hat{\Pi}_{b|y} &&\forall\; b,y\\ 
        & \sum_a Z_a=\mathbb{1}, \quad Z_a\geq 0 &&\forall\; a\\
        & \;[\hat{\Pi}_{a|x},\hat{\Pi}_{b|y}]=[\hat{\Pi}_{a|x},Z_{c}]=[Z_c,\hat{\Pi}_{b|y}]=0 &&\forall\; a,b,c,x,y
        \end{aligned}
    \end{equation}
    from which the min-entropy $H_{\text{min}}(A|E)$ is computed as in Eq.~\eqref{Eq:min_entrpy}. In our case, we employed \texttt{Mathematica} to write the SDP hierarchy and define the constraints, which was later solved in \texttt{Matlab} using \texttt{YALMIP}~\cite{lofberg_yalmip_2004} and \texttt{Mosek} as a solver~\cite{mosek}.
    
    \item The previous SDP optimization, allows us to construct a Bell operator of the form
    \begin{equation}
        \hat{\mathcal{B}}
            = \sum_{x,y,a,b}
                c_{x,y,a,b} 
                \hat{\Pi}_{a\vert x}(\boldsymbol{\theta}_x)
                    \otimes
                \hat{\Pi}_{b\vert y}(\boldsymbol{\theta}_y),
    \end{equation}
    where the coefficients $c_{x,y,a,b}$  are obtained from the dual of our SDP problem. We denote the eigenvalues and eigenvectors of these Bell operator as $\lambda(\boldsymbol{\theta})$ and $\ket{\varphi(\theta)}$, with $\boldsymbol{\theta} = \{\boldsymbol{\theta}_{x},\boldsymbol{\theta}_{x}\}_{x,y}$. From these, one can define updated versions of Alice and Bob's measurement settings by searching the optimal eigenvalues of this Bell operator, that is
    \begin{equation}\label{Eq:eigenval:op}
        \boldsymbol{\theta}^{(1)}
            = \underset{{\boldsymbol{\theta}}}{\text{optimize}}
                \big[
                    \lambda(\boldsymbol{\theta})
                \big],
    \end{equation}
    with the state shared by Alice and Bob being updated as $\ket{\psi_1} = \ket{\varphi(\boldsymbol{\theta}^{(1)})}$. This optimization is performed using local-optimization-based methods. Specifically, to evaluate Eq.~\eqref{Eq:eigenval:op}, we employed the \texttt{MATLAB Multistart} algorithm from the \texttt{Global Optimization} package which, in brief, launches several points and keeps the optimal one~\cite{MatlabGO}. It is worth noting that, one of the initial points for this algorithm was set to be equal to the $\boldsymbol{\theta}^{(0)}$ used in step 1.
    

    \item Steps 1-3 are iterated such that, at the $i$th step, we obtain the parameters $\boldsymbol{\theta}_{i}$ and the state $\ket{\psi_i}$ leading an optimized version of the min-entropy $H^{(i)}_{\text{min}}$. The iteration between these steps is performed until the condition $\lvert H_{\text{min}}^{(i)}-H_{\text{min}}^{(i-1)}\rvert < \epsilon$ is satisfied. In practice, $\epsilon$ was chosen to be $10^{-4}$, which is approximately two orders of magnitude above the precision of the algorithm used for the SDP optimization. Then, these parameters are used for estimating the von-Neumann entropy using the method in Ref.~\cite{brown2023deviceindependent} when setting $M=16$ (for more details see SM \ref{App:BFF}). This has been done in \texttt{Python} using the \texttt{nscpol2sdpa} package~\cite{wittek_algorithm_2015}, more specifically the update introduced in Ref.~\cite{PBrown_Github}, using the Python version of \texttt{Mosek}~\cite{mosekpython}.

    \item Once this convergence condition is met, we optimize the extra measurement setting used by Bob for constructing the key rate, which we denote as $\boldsymbol{\theta}_{y^*}$, such that $H(A\vert B)$ becomes minimum. Similarly to step 3, these consist of a local optimization method analogous to that descrbed in step 3.
    We denote the out-coming relative entropy in this cases as $H_{\text{opt}}(A\vert B)$.

    \item Finally, from the values obtained in steps~4 and 5, we compute the optimal value of the key rate as
    \begin{equation}
        r_{\text{opt}}
            = H^{(M)}(A\vert E)
                - H_{\text{opt}}(A\vert B).
    \end{equation}
\end{enumerate}

\begin{figure}
    \centering
    \includegraphics[width = 0.8\textwidth]{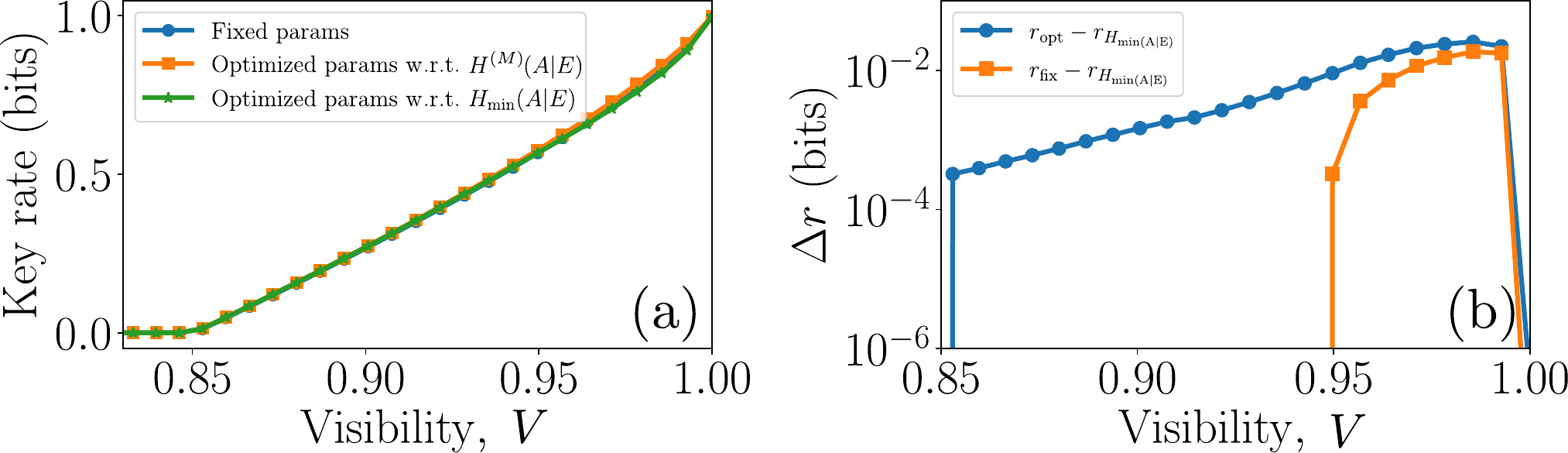}
    \caption{In (a), the key rate for the $d=2, m=2$ scenario is depicted as a function of visibility across various methodologies. Specifically, it is computed using fixed parameters from Ref.~\cite{salavrakosBellInequalitiesTailored2017} (blue curve with circular markers), direct optimization of $H^{(M)}(A|E)$ following the method detailed in the Supplementary Material of Ref.~\cite{gonzalez-ruiz_device_2022} (orange curve with squared markers), and the approach outlined based on the optimization shown in Fig.~\ref{fig:scheme:optimization} (green curve with star markers). In (b), the difference among these key rates is depicted against the visibility.}
    \label{Fig:diff:d2m2}
\end{figure}

In Fig.~\ref{Fig:diff:d2m2}~(a),we present the key rate obtained for the $d=2,m=2$ case. This includes results obtained with fixed parameters as outlined in Ref.~\cite{salavrakosBellInequalitiesTailored2017} (blue curve with circular markers), results derived through the direct optimization of $H^{(M)}(A|E)$ following the methodology detailed in the Supplementary Material of Ref.~\cite{gonzalez-ruiz_device_2022} (orange curve with squared markers), and the proposed method (green curve with star markers). In this scenario, all approaches yield highly comparable outcomes. However, a more detailed analysis highlighting their discrepancies is presented in Fig.\ref{Fig:diff:d2m2}(b). Here, we observe that the method demonstrated in the Supplementary Material of Ref.~\cite{gonzalez-ruiz_device_2022} (blue curve with circular dots) yields optimal rates, particularly evident as the critical visibility is approached. Conversely, comparing our proposed method with the fixed parameter case (orange curve with squared markers), it becomes apparent that our approach is suboptimal in scenarios with high visibilities, although it showcases improved performance as visibility decreases.

\begin{figure}[h!]
    \centering
    \includegraphics[width = 0.8\textwidth]{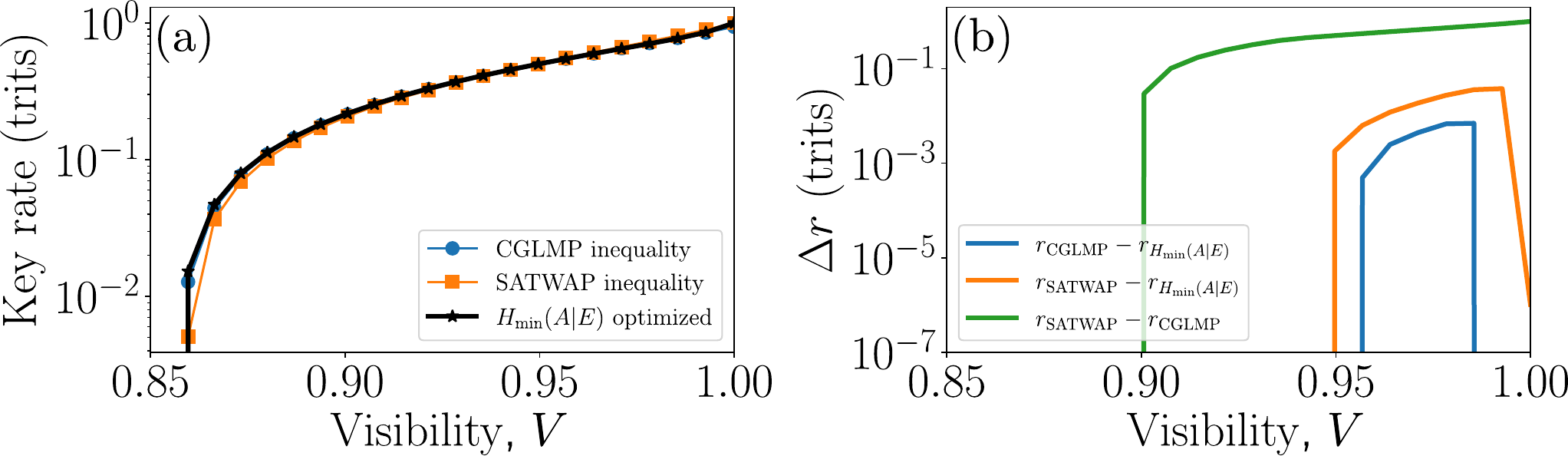}
    \caption{In (a), the key rate for the $d=3, m=2$ scenario is depicted as a function of visibility across different methodologies. Specifically, it is computed using the CGLMP inequality (blue curve with circular markers), the maximal violation concerning Salavrakos' inequality (orange curve with squared markers), and the method outlined here based on the optimization shown in Fig.~\ref{fig:scheme:optimization} (green curve with star markers). In (b), the disparity among these key rates is illustrated as a function of visibility.}
    \label{Fig:diff:d3m2}
\end{figure}

In Fig.\ref{Fig:diff:d3m2}~(a), we present the key rate obtained for the $d=3, m=2$ scenario computed using the maximal violation achieved for the CGLMP inequality (blue curve with circular markers) and the maximal violation of the Salavrakos' inequality (orange curve with squared markers), alongside the optimization method described herein. Notably, unlike the $d=2, m=2$ case where optimizing $H^{(M)}(A|E)$ was feasible using the methodology detailed in the Supplementary Material of Ref.~\cite{gonzalez-ruiz_device_2022}, it becomes impractical now due to the local optimization methods necessitating numerous evaluations of the SDP problem associated with $H^{(M)}(A|E)$. A single evaluation already requires several hours. Nevertheless, leveraging optimization techniques based on min-entropy, we observe that the results obtained offer superior bounds for the key rate, particularly nearing critical visibility. However, as demonstrated in Fig.\ref{Fig:diff:d3m2}~(b), this approach provides suboptimal values for the key rate in the high-visibility region. Nonetheless, this discrepancy is relatively modest, in the range of $10^{-2}-10^{-3}$ when compared to utilizing the two aforementioned inequalities.


\section{Relaxations on the von Neumann entropy}\label{App:BFF}
To lower bound the von Neumann entropy $H(A|x=x^*, y=y^*, E)$, we followed the method presented in Ref.~\cite{brown2023deviceindependent}. In this case, the convex-optimization problem we solved reads as

\begin{equation}
    \begin{aligned}\label{BFF_SDP} 
    \tilde{H}^{(M)}(A\vert x=x^*, y=y^*, E) = c_M+ \sum_{i=1}^{M-1} \frac{w_i}{t_i \ln d}\; \text{inf}\;& \sum_a \text{Tr}\big[ \hat{\rho}_{ABE}\big(\hat{\Pi}_{a|x^*}(Z_{a,i}+Z^*_{a,i}+(1-t_i)Z^*_{a,i}Z_{a,i})+t_i Z_{a,i} Z^*_{a,i}\big)\big]\\
     \text{s.t.} & \;\text{Tr} \big[ \hat{\rho}_{ABE}(\hat{\Pi}_{a|x}\hat{\Pi}_{b|y} )\big]=p(a,b|x,y)\\ 
        &\sum_a \hat{\Pi}_{a|x}=\sum_b \hat{\Pi}_{b|y}=\mathbb{1} \hspace{3.45cm}\forall\; x,y\\ 
        &\;\hat{\Pi}_{a|x}\geq 0, \quad \hat{\Pi}_{b|y}\geq 0 \hspace{3.905cm}\forall\; a,b,x,y\\ 
        &\;\hat{\Pi}^2_{a|x}=\hat{\Pi}_{a|x} \hspace{5.2cm} \forall\; a,x\\ 
        &\;\hat{\Pi}^2_{b|y}=\hat{\Pi}_{b|y} \hspace{5.29cm} \forall\; b,y\\ 
        &\;[\hat{\Pi}_{a|x},\hat{\Pi}_{b|y}]=[\hat{\Pi}_{a|x},Z^{(*)}_{c,i}]=[Z^{(*)}_c,\hat{\Pi}_{b|y}]=0 \hspace{0.5cm}\forall\; a,b,c,x,y
    \end{aligned}
\end{equation}
where $d$ is the dimension of the system, $M\in\mathbb{N}$, $w_i$ and $t_i$ are the wights and nodes of a $M$-point Gauss-Radau quadrature with $t_M=1$ and $c_M=\sum_{i=1}^{M-1}\frac{w_i}{t_i \ln d}$.
To compute $\{p(ab|xy)\}$ we considered the state and measurements obtained by the optimization in step 4 of the procedure explained in SM \ref{App:Optimization}. 

It is worth noting that in the definition of these SDP problems, Eve's operators $Z_a$ are no longer elements of a POVM. In general they could also be non-hermitian. Furthermore, better versions of this bound can be obtained by moving the infimum out from the summation over $a$ in Eq.\eqref{BFF_SDP}. More explicitly, instead of optimizing
\begin{equation}
    \tilde{H}^{(M)}(A\vert x=x^*, y=y^*, E) = c_M + \sum_{i=1}^{M-1} \frac{w_i}{t_i \ln d}\; \text{inf}\; \sum_a \text{Tr}[\dots],
\end{equation}
one could consider
\begin{equation}
    H^{(M)}(A\vert x=x^*, y=y^*, E) = c_M + \text{inf}\;\sum_{i=1}^{M-1} \frac{w_i}{t_i \ln d} \sum_a \text{Tr}[\dots],
\end{equation} 
which is proven to converge to the Von Neumann entropy $H(A \vert x=x^*, y=y^*, E)$ when $M\rightarrow \infty$. In general, the following inequality chain holds $H(A \vert E) \geq  H^{(M)}(A \vert E) >  \tilde{H}^{(M)}(A \vert E)$,  for every value of $M$.
Optimizing $H^{(M)}(A \vert E)$, would require solving a single but excessively large SDP, which would drastically increase the computation time.


\section{Analytical derivation of the upper bound on the key rate using the maximally entangled state} \label{appendix:analytical_derivation}

The explicit form of the CGLMP Bell expression from Collins \emph{et al.} \citep{collinsBellInequalitiesArbitrarily2002} is
    \begin{equation} \label{eq:CGLMP_meas}
        \begin{aligned}
        I_d^{x_1,x_2,y_1,y_2} &= \sum_{k=0}^{[d / 2]-1}\left(1-\frac{2 k}{d-1}\right) \{ p(A_{x_1}=B_{y_1}+ k) \\
        & +p(B_{y_1}=A_{x_2}+k+1 )+p(A_{x_2}=B_{y_2}+k) \\
        &+p(B_{y_2}=A_{x_1}+k) -p(A_{x_1}=B_{y_1}-k-1) \\
        &-p(B_{y_1}=A_{x_2}-k) -p(A_{x_2}=B_{y_2}-k-1) \\
        & -p(B_{y_2}=A_{x_1}-k-1) \}
        \end{aligned}.
    \end{equation}
where $p(A_x=B_y+k)$ is the probability that Alice's and Bob's outcomes differ by $k$ modulo $d$ for measurement settings $x$ and $y$. That is
\begin{equation} \label{eq:probs_k}
    p(A_x=B_y+k) \coloneqq \sum_{j=1}^{d} p_{AB}(j,j+k \textup{ mod }d |x,y).
\end{equation}
For local variable theories, $I_d \le C_b \coloneqq 2$. Since the measurement settings are irrelevant, we have omitted the superindices $x_1,x_2,y_1,y_2$ in $I_d$.

We use the CGLMP-optimal measurements for settings $x,y\in\{1,2\}$. These maximize the value of $I_d^{1,2,1,2}$ achieved by the maximally entangled state $\ket*{\psi_+} = \frac{1}{\sqrt{d}}\sum_{q=1}^{d} \ket*{qq}$.  This maximal value is \citep{collinsBellInequalitiesArbitrarily2002}
\begin{equation} \label{eq:Idp}
I_d^\textup{max} = 4 d \sum_{k=0}^{[d / 2]-1}\left(1-\frac{2 k}{d-1}\right)\left(f_d(k)-f_d(-(k+1))\right),    
\end{equation}
where $f_d(k) \coloneqq 1/(2d^3\sin^2 [ \pi(k+1/4)/d ])$. Using Eqs. \eqref{eq:prob_loc} and \eqref{eq:probs_k}, we find that
\begin{equation}
        p^\mathcal{L} (A_x = B_y + k) = \frac{1-\tilde{V}}{d} + \tilde{V} p^\mathcal{NL} (A_x = B_y + k).
\end{equation}
By substituting this into Eq.~\eqref{eq:CGLMP_meas}, we get $I_{d}^{\mathcal{L}} = \tilde{V} I_{d}^{\mathcal{NL}}$. Similarly, we can see that $I_{d}^\textup{obs} = V I_{d}^{\mathcal{NL}}$. By setting $I_d^\mathcal{L} = C_b$ in the first expression, we get $\tilde{V} = C_b/I_d^\textup{max}$. Hence, if $p_{AB}^\mathcal{L}$ is local with this value of $\tilde{V}$, then $\tilde{V}$ must be maximal, since $p_{AB}^\mathcal{L}$ reaches the local bound of the CGLMP-inequality, and therefore any larger value of $\tilde{V}$ would imply a Bell inequality violation. We can verify that $p_{AB}^\mathcal{L}$ is local by checking that if we set $\tilde{V} = C_b/I_d^\textup{max}$ in Eq.~\eqref{eq:prob_loc}, then the resulting probability distribution can be decomposed as a convex-combination of deterministic strategies. We do this via linear programming up to $d=10$. We conjecture that this is the case for any $d$, and hence $V^\mathcal{L} = C_b/I_d^\textup{max}$ $\forall d \ge2$.

Note that, when using the maximally entangled state, the probabilities $p_{AB}^\mathcal{NL}(a,b|x,y)$ only depend on the differences between the outcomes $a$ and $b$ modulo $d$, and on certain parameters characterizing the measurements performed by Alice and Bob \citep{salavrakosBellInequalitiesTailored2017}. If Alice and Bob use the same measurement parameters for the key settings, then $p_{AB}^\mathcal{NL}(a,b|x^*,y^*) = \delta_{a,b}/d$.

Next, we can calculate the upper bound on the key rate. Recall that $r_\textup{ub} \coloneqq H(A|E)-H(A|B)$, where $H(A|E)$ is the PA-term and $H(A|B)$ is the EC-term. The conditional entropy of $Y$ given $X$ is $H(Y|X) = \sum_{x\in\mathcal{X}} p(x) H(Y|X=x)$, where $H(X) = -\sum_{x\in\mathcal{X}} p(x) \log_d p(x)$ is the Shannon entropy. With this in mind, the EC-term can be written as
\begin{equation}
    H(A|B) = \sum_{b=1}^{d} p_B^\textup{obs}(b) H\left\{ \frac{p_{AB}(1,b)}{p_B(b)},\dots,\frac{p_{AB}(d,b)}{p_B(b)} \right\}
\end{equation}
where all probabilities are for the key settings. Since $p_{AB}(a,b) = V p_{AB}^\mathcal{NL}(a,b) + (1-V)/d^2$ only depends on the difference between the outcomes $a$ and $b$ modulo $d$, and since $p_B^\textup{obs}(b)=1/d$ $\forall b$,
\begin{equation}
    \begin{aligned}
        H(A|B) &= H\left\{ V + \frac{1-V}{d}, \frac{1-V}{d}, \dots, \frac{1-V}{d}  \right\} \\
        &= - \frac{1+(d-1)V}{d}\log_d \left(1+(d-1)V \right) - \frac{(d-1)(1-V)}{d} \log_d\left( 1-V \right) + 1.
    \end{aligned}
\end{equation}

For the PA-term, since Eve has perfect knowledge of Alice's outcomes in the local rounds and has no knowledge of the outcomes of the non-local rounds, we have $H(A|E,\mathcal{L})=0$ and $H(A|E,\mathcal{NL})= 1$. Hence, $H(A|E) = q^\mathcal{NL} = 1-q^\mathcal{L}$. Using Eq.~\eqref{eq:qL} and the fact that $V^\mathcal{L} = C_b/I_d^\textup{max} =  2/I_d^\textup{max}$, we get
\begin{equation}
    H(A|E) = 1- \frac{1-V}{1-2/I_d^\textup{max}}
\end{equation}
if $V\ge V^\mathcal{L}$ and $H(A|E) = 0$ otherwise.

\section{Additional tables and figures} \label{sec:additional_figures}

\begin{table}[h!]
    \centering
    \begin{tabular}{|c|c|c|} 
    \hline
    \multicolumn{1}{|c|}{} & \multicolumn{2}{c|}{$V_\textup{crit}$} \\ \hline
    $d$ & Maximally entangled & CGLMP \\ \hline
    2 & 0.82999 & 0.82999 \\
    3 & 0.82043 & 0.82101 \\
    4 & 0.81464 & 0.81550 \\
    5 & 0.81064 & 0.81165 \\ 
    6 & 0.80766 & 0.80874 \\
    7 & 0.80532 & 0.80644 \\
    8 & 0.80341 & 0.80455 \\ 
    \hline
    \end{tabular}
    \caption{Critical visibilities for dimensions ranging from two to eight when using a mixture of local deterministic strategies and the maximally entangled state or the CGLMP state.}
    \label{tab:Vcrit}
\end{table}

\begin{figure}[h]
    \centering
    \includegraphics[width=0.8\textwidth]{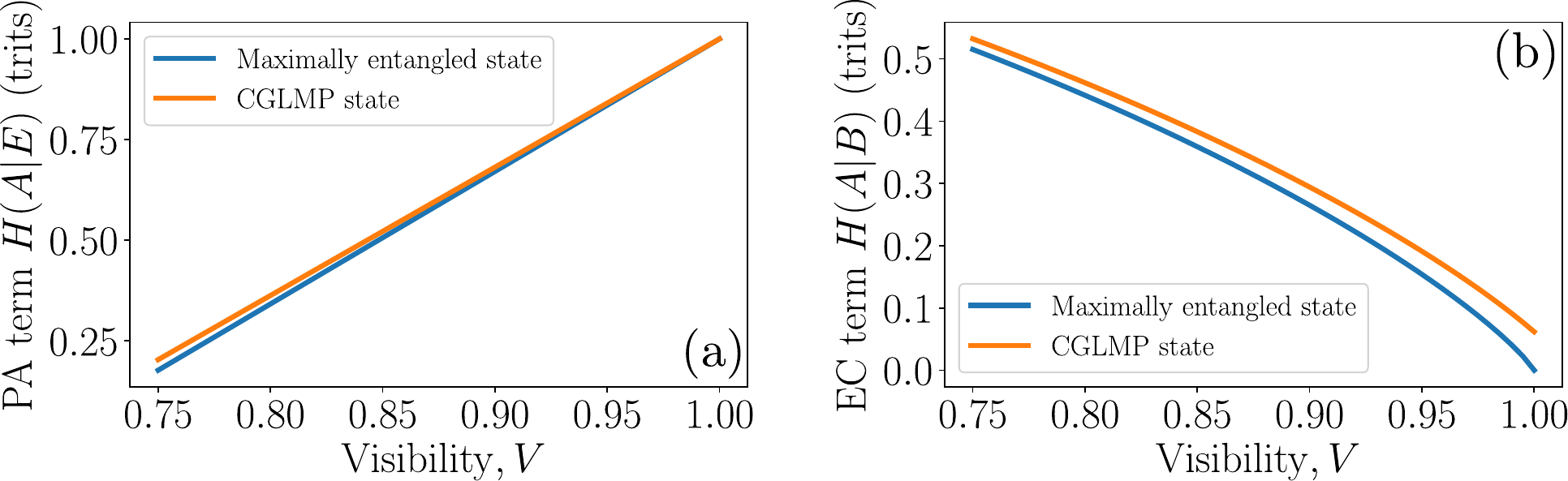}
    \caption{In (a), the PA-term in the CC-based upper bound on the key rate is presented as a function of the visibility when using the maximally entangled state, and when using the CGLMP state for dimension $d=3$. For $V$ close to one, the values are very close to each other in both cases. In (b), the dependence of the EC-term in the CC-based upper bound with respect to the key rate in terms of the visibility when using the maximally entangled state, and when using the CGLMP state for dimension $d=3$. For $V$ close to one, the value of the EC-term is significantly larger when using the CGLMP state as opposed to the maximally entangled state. This is due to the fact that the outcomes will be maximally correlated when using the maximally entangled state, and will therefore require the least amount of error correction. This also explains why the CGLMP state has a slightly larger critical visibility for DIQKD.}
    \label{fig:PA:EC_vs_d}
\end{figure}


\end{document}